\documentclass[11pt,a4paper]{article}
\pdfoutput=1
\usepackage{amsmath,amssymb,graphicx,subfig,bm,euscript,cancel,mathtools,setspace,color,mathrsfs,hyperref}
\usepackage[utf8]{inputenc}
\usepackage[sort&compress,square,numbers]{natbib}
\usepackage[left=35mm,right=35mm,top=35mm,bottom=40mm]{geometry}
\numberwithin{equation}{section}
 \def\eq#1{Eq.~(\ref{#1})}
 
 \def\fig#1{fig.~{\ref{#1}}}
 \def\Fig#1{Fig.~{\ref{#1}}}
 \def\beq{\begin{equation}}
 \def\eeq{\end{equation}}
 \def\beqa{\begin{eqnarray}}
 \def\eeqa{\end{eqnarray}}
 
 \newcommand{\bs}{\boldsymbol}
 \def\one{\!\!{\hbox{ 1\kern-.8mm l}}}
 \newcommand{\bra}[1]{\langle{#1}|}
 \newcommand{\ket}[1]{|{#1}\rangle}
 
 \newcommand{\ex}[1]{{\rm e}^{#1}}
 \renewcommand{\d}{{\rm d}}
 \def\ii{{\rm i}}

\makeatletter
\newcommand{\dotminus}{\mathbin{\text{\@dotminus}}}
\newcommand{\@dotminus}{%
  \ooalign{\hidewidth\raise1ex\hbox{.}\hidewidth\cr$\m@th-$\cr}%
}
\makeatother
\newcommand{\dm}{\dotminus}

\newcommand{\tran}{^{\text{t}}}

\DeclareMathAlphabet\EuRoman{U}{eur}{m}{n}
\SetMathAlphabet\EuRoman{bold}{U}{eur}{b}{n}

\newcommand{\ie}{\emph{i.e.}}
\newcommand{\eg}{\emph{e.g.}}
\newcommand{\mb}[1]{\mathbb{#1}}
\newcommand{\psl}{\ensuremath{\text{PSL}(2)}}
\newcommand{\slc}{\ensuremath{\text{SL}(2)}}

\newcommand{\rs}{\ensuremath{\mb{CP}^1}}
\newcommand{\surf}{\ensuremath{\Sigma}}
\newcommand{\sg}{\ensuremath{G}}
\newcommand{\wt}{\ensuremath{\widetilde}}
\newcommand{\wh}{\ensuremath{\widehat}}
\newcommand{\re}{_{\text{red}}}
\newcommand{\osp}{\ensuremath{\text{OSp}(1|2)}}
\newcommand{\ve}{\varepsilon}
\newcommand{\cf}{\emph{cf.}}
\newcommand{\hyp}{\ensuremath{\text{Hyp}(2|1)}}
\newcommand{\tz}{\ensuremath{{\wt{z}}}}
\newcommand{\tw}{\ensuremath{{\wt{w}}}}

\newcommand{\cc}{\ensuremath{\check{c}}}
\newcommand{\ber}{\text{Ber}}

\newcommand{\sscsp}{\ensuremath{\bs{\frak{S}}_g}}
\newcommand{\srs}{\ensuremath{\bs{\surf}}}
\newcommand{\PT}{\ensuremath{\mathit{\Pi T}}}
\allowdisplaybreaks
\begin{document}
\begin{titlepage}
\begin{flushright}
\vspace*{-25pt}
\end{flushright}
\begin{center}
\vspace{1cm}
{\Large \bf Deforming super Riemann surfaces with gravitinos and super Schottky groups}

\vspace{8mm}
Sam Playle
\footnote{email: {\tt playle@to.infn.it}}
\vskip .5cm
{\sl Dipartimento di Fisica, Universit\`a di Torino  \\
and INFN, Sezione di Torino}\\
{\sl Via P. Giuria 1, I-10125 Torino, Italy}\\
\vskip 1.2cm
\begin{abstract}
\noindent
The (super) Schottky uniformization of compact (super) Riemann surfaces is briefly reviewed. Deformations of super Riemann surface by gravitinos and Beltrami parameters are recast in terms of super Schottky group cohomology. It is checked that the super Schottky group formula for the period matrix of a non-split surface matches its expression in terms of a gravitino and Beltrami parameter on a split surface. The relationship between (super) Schottky groups and the construction of surfaces by gluing pairs of punctures is discussed in an appendix.
\end{abstract}
\end{center}

\vfill

\end{titlepage}
\tableofcontents
\section{Introduction}
\label{intro}
There has been a resurgence of interest in super Riemann surfaces (SRS) and their supermoduli \cite{Witten:2012ga,Donagi:2013dua,Donagi:2014hza,Jost:2014wfa,Witten:2015hwa,D'Hoker:2015kwa}. SRS are the natural objects on which to define superconformal field theories in two dimensions \cite{Friedan:1986rx}, and thus they play a fundamental role in perturbative superstring theory in the Ramond-Neveu-Schwarz framework, where quantum amplitudes are computed as integrals of certain measures over supermoduli space \cite{Friedan:1985ge}. For a classical review of how these measures arise, see, for example, section III of \cite{D'Hoker:1988ta}, or \cite{Witten:2012bh,Witten:2013cia,Witten:2013tpa} for a more recent exposition. An understanding of these objects is thus important in theoretical physics as well as for the considerable pure-mathematical appeal they hold.

There are several ways to compute with super Riemann surfaces used in the physics literature. The most common is to work with the underlying Riemann surface along with a spinor-valued differential form called a gravitino (a recent work demonstrating the validity of this description is \cite{Jost:2014wfa}). This is the approach taken, for example, by D'Hoker and Phong to compute the period matrix of a non-split SRS  \cite{D'Hoker:1989ai}, which was used to fibre supermoduli space over moduli space for genus $g=2$ and to thus evaluate the two-loop superstring vacuum energy (see, for example, \cite{D'Hoker:2014nna} and references therein).

A different approach is that of super Schottky groups \cite{Martinec:1986bq,Manin1986,DiVecchia:1989id}, with which compact SRS are given as quotients of a covering space by certain groups $\bs G \subseteq \osp$ of `super-projective' maps. The construction is directly analogous to the classical construction of compact Riemann surfaces as quotients by Schottky groups \cite{Ford1929}. Super Schottky groups have some drawbacks: they treat $A^i$ and $B_j$ homology cycles on a different footing (in the conventional basis), it is not known how to explicitly characterize all boundaries of super-Schottky space, and they describe only those SRS with even spin structures.

But super Schottky groups have a number of attractive features: automorphic forms can be used to give explicit formulae for many objects (such as the period matrix $\bs\tau_{ij}$, the abelian differentials, the Szeg\H o kernel, the prime form, and so on). The supermoduli are realized fairly explicitly as even and odd parameters of a set of super-projective transformations which generate the group. Super Schottky space has a natural complex structure compatible with supermoduli space (in contrast to the uniformization by super-Fuchsian subgroups of \osp~\cite{Crane:1986uf,Manin1986}). It is appealing to think of super Schottky groups as arising from the repeated gluing of pairs of marked points (\ie~Neveu-Schwarz punctures) on SRS, giving the supermoduli an intuitive interpretation, and making them well-suited to the description of certain types of degeneration.

Because of their relationship to gluing, (bosonic) Schottky groups emerged automatically in the earliest attempts at computing multiloop string theory amplitudes with operator methods \cite{Lovelace:1970sj,Kaku:1970ym,Alessandrini:1971cz,Alessandrini:1971dd}, with super Schottky groups emerging from analogous superstring computations in the 1980s \cite{DiVecchia:1989id}. The fact that Schottky moduli can be easily related to gluing parameters near corners of moduli space means that Schottky groups are particularly well-suited to describing the low-energy behaviour of string theory amplitudes. For example, in \cite{Magnea:2013lna}, complicated two-loop Feynman integrands for Yang-Mills gauge theory amplitudes have been reproduced starting from the bosonic string measure described by Schottky groups, as expected. That result has been reproduced successfully also starting from the Neveu-Schwarz sector of Type IIB superstring theory using super Schottky groups in \cite{Magnea:2013lna,Magnea:2015fsa}, with the benefit that there is no longer a tachyon contribution as in the bosonic theory.

This paper is concerned with the relationship between the super Schottky uniformization and the description of non-split SRS by gravitinos (and metric deformations) on split surfaces. In particular, we want to translate results expressed using gravitinos into a form which is useful for super Schottky group computations. The main idea is to take statements involving the deformation of split SRS by gravitinos and Beltrami differentials and recast them in terms of the Eichler cohomology of the super Schottky group. Our primary goal is to check that the formula for the period matrix of a non-split SRS given in \cite{D'Hoker:1989ai} and section 8 of \cite{Witten:2012ga} gives the same result as the super Schottky group series formula given in \cite{DiVecchia:1989id}.

Our calculation uses first-order deformations which can be described with quasisuperconformal vector fields \cite{Martinec:1986bq}. By taking the deformation to be a nilpotent function of two odd parameters we can ensure that it is identical to its linear approximation. Restricting to two odd supermoduli is completely general in genus two (with no punctures) when this matches the odd dimension of supermoduli space. We check equalities by computing the first few terms in power series expansions in the (super) Schottky (semi)multipliers. Since these moduli can be thought of as gluing parameters describing the pinching of the $A^i$ homology cycles, this expansion captures the leading behaviour near the corresponding corner of the Deligne-Mumford compactification of (super) moduli space.

The motivation of this work is to develop techniques suitable for adapting super Schottky groups to the Ramond sector of superstring theory.
Super Schottky groups are useful for describing string worldsheets near corners of moduli space only when the string states propagating through the nodes are in the Neveu-Schwarz sector, because super Schottky moduli are related to gluing parameters for pairs of Neveu-Schwarz punctures. Super Schottky groups are not suitable for describing string worldsheets near corners of moduli space with Ramond nodes. This precludes, in particular, the use of super Schottky groups to compute integrands for Feynman graphs with fermion edges by generalizing the results of \cite{Magnea:2015fsa}. A generalization of super Schottky groups allowing both spin structures for the $A^i$ cycles would thus be considerably useful (some progress in this direction was made by Petersen in \cite{Petersen:1989hi}).

The outline of this paper is as follows. In section \ref{sgsect}, we review necessary facts about the (super) Schottky group construction of compact (super) Riemann surfaces, also recalling some points about super Riemann surfaces. Section \ref{defschot} contains the main results of this paper: in subsection \ref{bosdef} we collect results about deformations of Riemann surfaces via Beltrami differentials and how they can be related to shifts in the Schottky moduli, then in subsection \ref{ssgdef} we see how the analysis is adapted for deformations of super Riemann surfaces, using our results to compute the period matrix of a non-split super Riemann surface in genus $g=2$.

In Appendix \ref{schotapp} we include some (super) Schottky group formulae for geometric objects defined on (super) Riemann surfaces and (super) moduli space. In Appendix \ref{SewingApp} we discuss the relationship between (super) Schottky groups and the construction of higher-genus compact (super) Riemann surfaces by gluing plumbing fixtures between pairs of marked points, with the Schottky (super) moduli arising in a simple way.

\subsubsection*{Notation} We write `bosonic' to refer specifically to objects in non-super geometry. When we use `odd constants' \eg~odd super Schottky parameters $\theta_i$, $\phi_i$, we are implicitly working over a Grassmann algebra generated by these constants. A bold italic letter \eg~$\bs w, \bs z$ denotes a function valued in $\mb{CP}^{1|1}$ or $\mb{C}^{1|1}$ (possibly depending on odd constants), such as a superconformal coordinate or an element of \osp. To avoid clutter we do not distinguish notationally between linear and (super-)conformal realizations of \psl~and \osp. We use Greek indices for generic (super) Schottky group elements: $\gamma_\alpha$; $\bs \gamma_\beta$, with Roman indices specifically for the generators: $\gamma_i$; $\bs \gamma_j$.
\subsubsection*{\emph{Mathematica} notebook} A \emph{Mathematica} notebook, `\verb=GravitinoSchottky.nb=', and a package, `\verb=schottky.m=', are included as ancillary files on arXiv. In the notebook we check the results of sections \ref{permatsubs} and \ref{schoformgtwo}, in which the period matrix of a non-split SRS in genus $g=2$ is computed to first order in the semimultipliers $\ve_i$, both directly from the formula \eq{schopm} and via deformations of a split surface, \eq{taufromdef}.

\section{Schottky Groups}
\label{sgsect}
\subsection{Schottky groups for Riemann surfaces}
\label{schot}
In this section we review a classical construction of families of compact Riemann surfaces as quotients of a certain covering space by the action of a \emph{Schottky group} of M\"obius maps. Here we simply state the construction; in Appendix \ref{SewingApp} we explain how it can be arrived at by gluing pairs of marked points.
\subsubsection{Hyperbolic M\"obius maps}
Schottky groups are subgroups of \psl, the group of matrices of the form
\begin{align}
\gamma & \equiv \left(\begin{array}{cc} a & b \\ c & d \end{array}\right) \, , &  ad - bd & = 1 \, ,
\end{align}
subject to the equivalence relation $(\begin{smallmatrix}a & b \\ c & d \end{smallmatrix})\sim (\begin{smallmatrix}-a &- b \\ -c & -d \end{smallmatrix})$.
\psl~acts on the Riemann sphere $\rs = \mb{C} \cup \{ \infty \}$ by M\"obius maps:
\begin{align}
z  \mapsto \gamma(z) & = \frac{az+b}{cz + d} \, . \label{moebius}
\end{align}
A M\"obius map with two distinct fixed points, one attractive and one repulsive, is called \emph{hyperbolic} (or loxodromic).  Any hyperbolic M\"obius map $\gamma$ with attractive and repulsive fixed points $u$ and $v$, respectively, can be defined implicitly by \cite{Ford1929}
 \begin{align}
 \frac{\gamma(z) - u}{\gamma(z) - v} & = k\, \frac{z - u}{z - v} \, , \label{gamuvk}
 \end{align}
 where $k$ is called the \emph{multiplier} of $\gamma$, with $0<|k|<1$. Hyperbolic M\"obius maps are therefore parametrized by $u$, $v$ and $k$.
Explicitly, we can write
\begin{align}
\gamma & = \Gamma_{uv}^{-1} \circ P_k \circ \Gamma_{uv} \label{SgenGam}
\end{align}
 where $\Gamma_{uv}$ is a \psl~map taking the attractive and repulsive fixed points $u$ and $v$ to $0$ and $\infty$, respectively, for example:
\begin{align}
\Gamma_{uv} & \equiv \frac{1}{\sqrt{u-v}} \left( \begin{array}{cc} 1 & - u \\ 1 & - v \end{array}\right) \, ,
& z \mapsto \Gamma_{uv}(z) & = \frac{z-u}{z-v} \, , \label{Gamuvdef}
\end{align}
and $P_k$ is a dilatation:
 \begin{align}
P_k & \equiv \left( \begin{array}{cc} \sqrt{k} & 0 \\ 0 & 1 / \sqrt{k} \end{array}\right) \, ,
&
z   \mapsto P_k (z) & = k z  \, . \label{Pkdef}
 \end{align}
The matrix $\gamma$ has eigenvectors $(u,1)\tran$ and $(v,1)\tran$ and the ratio of the corresponding eigenvalues is $k$.
\subsubsection{Schottky groups}
\label{schothigherg}
 \begin{figure}
\centering
\def\svgwidth{6cm}
\subfloat[]{ 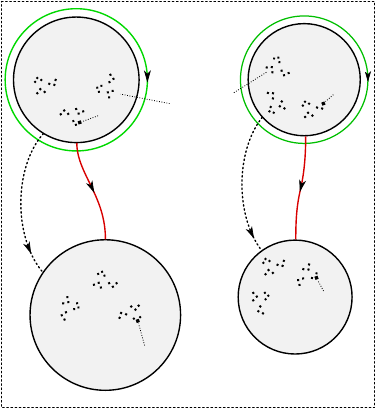 \label{fig:ssa} }
\def\svgwidth{6cm}
\subfloat[]{ 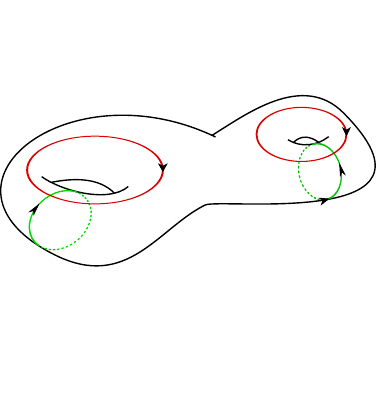 \label{fig:ssb} }
\caption{Quotienting $\rs - \Lambda(G)$ (\Fig{fig:ssa}) by a rank-2 Schottky group $G$ freely generated by $\{ \gamma_1, \gamma_2\}$ adds two handles to the sphere, giving a compact surface $\surf$ of genus $g=2$ (\Fig{fig:ssb}) conformally equivalent to the fundamental region $\overline{\cal F}(G)$ with its boundary circles identified in pairs. The standard basis of $A^i$ and $B_j$ homology 1-cycles is shown.}\label{schothandlesb}
\end{figure}
Suppose we have $2g$ circles on the Riemann sphere, say ${\cal C}_i$, ${\cal C}_i'$ for $i=1,\ldots,g$, which together bound a connected region ${\cal F}$ such that
\begin{align}
 \partial {\cal F} & = \sum_{i=1}^g({\cal C}_i'-{\cal C}_i) \label{Fbound}
 \end{align}
 where the sign denotes orientation. Furthermore, suppose we can find a set of $g$ hyperbolic M\"obius maps $\gamma_i \in \psl$ such that  $\gamma_i({\cal C}_i) = {\cal C}_i'$.

Then the group $G$ freely generated by the $\gamma_i$ is a \emph{Schottky group} of genus $g$.
Every element $\gamma$ of a Schottky group $G$ is hyperbolic (and in fact every freely-generated subgroub of \psl~with this property is a Schottky group \cite{Maskit1967}).

$\cal F$ is a \emph{fundamental region} for $G$; we can write $\overline{\cal F}$ for its closure which includes the $2g$ `Schottky circles' ${\cal C}_i$, ${\cal C}_i'$. The group has a \emph{limit set} $\Lambda(G) \subseteq \rs$, which is the set of points which are not equivalent by the Schottky group $G$ to some point in $\overline{\cal F}$ (this does not depend on the choice of ${\cal F}$: we may alternatively define $\Lambda(G)$ as the set of accumulation points of the orbits of $G$). Then if we subtract the limit set from the Riemann sphere and quotient by $G$, the coset space is a compact Riemann surface of genus $g$:
\begin{align}
\surf_g & = \big( \rs - \Lambda(G)) / G \, .
\end{align}
 We could define the same surface perhaps more intuitively by taking the fundamental region $\overline{\cal F}$ and identifying the boundary circles in pairs with $z \sim \gamma_i(z)$ for $z \in {\cal C}_i$ and $\gamma_i(z) \in {\cal C}_i'$, making sure to note that the resulting surface depends, of course, {only} on the Schottky group $G$ and not on the choice of ${\cal F}$. See \Fig{schothandlesb} for an illustration of the genus $g=2$ case.

 A \emph{marked} Schottky group is a Schottky group $G$ with a choice of $g$ generators $\gamma_i$. Since fixing the generators fixes the whole group, of course, we can parametrize marked Schottky groups by giving the $3g$ parameters $u_i, v_i, k_i$, $i=1,\ldots,g$. Two Schottky groups conjugate by a \psl~transformation will describe the same Riemann surface $\surf_g$; to get rid of this redundancy we can always fix coordinates on \rs~with
 \begin{align}
 u_1 &  = 0 \, , & v_1 & = \infty \, , & v_2 & = 1 \, . \label{normeq}
 \end{align}
 A marked Schottky group satisfying \eq{normeq} is \emph{normalized}. The $(3g-3)$-dimensional space of marked, normalized Schottky groups of genus $g$ is called \emph{Schottky space} $\mathfrak{S}_g$.

 A  compact Riemann surface $\surf_g$ is \emph{marked} if it has a basis $\{A^i,B_j; \, i,j = 1,\ldots g\}$ of 1-cycles whose oriented intersection number is given by $(A^i, A^j) = (B_i, B_j) = 0$ and $(A^i , B_j ) = - (B_j , A^i ) = \delta^i_j$. Conventionally, the $A^i$ cycles on a Riemann surface $\surf_g$ given by a Schottky group $G$ may be taken as the Schottky circles ${\cal C}_i$ for some choice of ${\cal F}$, while the ${ B}_i$ cycles are given by a choice of curve connecting a point on ${\cal C}_i$ to the $G$-equivalent point on ${\cal C}_i'$.

It is a classical theorem that any compact Riemann surface $\surf_g$ can be constructed with a Schottky group $G$. It is not always possible to find ${\cal F}$ such that the boundary components are geometric circles; it is only necessary that they are closed curves with the right topology.

\subsection{Schottky groups for super Riemann surfaces}
There is an analogous construction to the one in the previous section which can be used to describe super Riemann surfaces (SRS). In section \ref{srs} we recall some basic definitions and facts about SRS, in sections \ref{osp} and \ref{hyp21} we review super-projective transformations, and in section \ref{ssg} we describe how these can be used to build families of compact SRS.
\subsubsection{Super Riemann surfaces}
\label{srs}
Let us recall that an SRS $\srs$ is a $1|1$-dimensional complex supermanifold with some additional structure, namely that its tangent bundle comes with a rank-$0|1$ sub-bundle ${\cal D} \subseteq T\surf$ such that any local non-zero section $D$ of ${\cal D}$ has the property that $D^2 \equiv \frac{1}{2} \{ D , D \}$ is nowhere proportional to $D$ \cite{Witten:2012ga}. It is always possible to choose local coordinates $\bs z \equiv z|\zeta$ where sections of ${\cal D}$ are proportional to
\begin{align}
D_\zeta & \equiv \partial_\zeta + \zeta \partial_z \, , \label{superD}
\end{align}
whose square $D_\zeta^2 = \partial_z$ is clearly linearly independent of $D_\zeta$. Coordinates with this property are called \emph{superconformal}. If $\wh{\bs{z}} = \wh z | \wh \zeta$ is a coordinate system which overlaps with the superconformal coordinate $\bs z$, then it is also superconformal if and only if
\begin{align}
D_\zeta \, \wh{z} & = \wh{\zeta} \, D_\zeta \, \wh{\zeta}  \, \label{superconf}
\end{align}
holds \cite{Friedan:1986rx}.
The superderivative transforms under the superconformal change of coordinates as
\begin{align}
D_\zeta & = F_{\wh{\bs{z}}} ( \bs z) D_{\wh \zeta} \, , &
F_{\wh{\bs{z}}}(\bs z) & \equiv D_\zeta\, \wh{\zeta}(\bs z) \, , \label{superjac}
\end{align}
where the `semijacobian' $F_{\wh{\bs{z}}}(\bs z)$ satisfies a chain rule
\begin{align}
F_{\bs w \circ \bs v}({\bs u}) & = F_{\bs w}({\bs v}(\bs u)) \cdot F_{{{\bs v}}}({\bs u}) \, .
\end{align}
Any supermanifold has a Berezinian bundle, and an SRS in particular has a Berezinian bundle generated by the symbol $[\d z | \d \zeta ] \equiv [ \d \bs z]$ which transforms oppositely to $D_\zeta$ under superconformal changes of coordinates \cite{Giddings:1987wn}:
 \begin{align}
 [\d z | \d \zeta ] & = F_{\wh{\bs z}}( \bs z)^{-1} [ \d \wh{z} | \d \wh{\zeta}]
 \end{align}
 so that $[\d \bs z] D_\zeta$ is coordinate-independent. In general, if $\phi(\bs z) [\d \bs z]^{h}$ is superconformally covariant, \ie~if $\phi(\bs z) = F_{\wh{\bs z}}(\bs z)^h \wh{\phi}(\wh{\bs z})$, then $\phi(\bs z)$ is a section of ${\cal D}^{h}$ called an $h/2$-superdifferential, or a superconformal primary of weight $h/2$.


Of particular interest are ${1}/{2}$-superdifferentials which can be identified with sections of the Berezinian bundle and inserted in contour integrals covariantly. On a compact SRS of genus $g$, there is a $(g|0)$-dimensional space of holomorphic $1/2$-superdifferentials (called abelian superdifferentials). We can write them in terms of local superconformal coordinates as $\wh\sigma(z|\zeta) = \wh \phi(z|\zeta) [ \d z | \d \zeta]$. Given a homology basis of $\srs_g$ we can fix a normalized basis $\{ \wh\sigma_i\}$ of abelian superdifferentials by
    \begin{align}
    \frac{1}{2\pi \ii} \oint_{A^i} \wh\phi_j(z|\zeta) [ \d z | \d \zeta] & = \delta^i_j \, , \label{supabnorm}
    \end{align}
    where the Berezinian integration is understood in the usual way (see, for example, \cite{Witten:2012bg}).
    Then the period matrix of $\srs_g$, $\bs \tau_{ij}$, is a symmetric $g \times g$ matrix defined by \cite{Voronov:1987xf}
    \begin{align}
    \bs\tau_{ij} & = \frac{1}{2\pi\ii} \oint_{B_j} \wh\phi_i(z|\zeta) [ \d z | \d \zeta] \, . \label{srspermatdef}
    \end{align}

As with any supermanifold, an SRS \srs~has an associated \emph{reduced} surface $\surf$ which is an ordinary Riemann surface: if \srs~has transition functions $\bs z_i(\bs z_j) = f_{ij} ( z_j | \zeta_j) \big| \phi_{ij} (z_j |\zeta_j)$, then the surface with transition functions $z_i(z_j) = f_{ij}(z_j|0)$ is the associated reduced surface $\Sigma$ \cite{Witten:2012ga}.

A simple example of a compact SRS is a `Riemann supersphere' \cite{Manin1986} defined by giving two superconformal charts $\bs z$ and $\bs w$ related by the superconformal transformation
\begin{align}
\bs{z} & = \bs{I} (\bs{w}) \, , &
\bs{I} (\bs{w}) & \equiv -1/w \big| - \psi / w \, . \label{ztoy}
\end{align}
As a complex supermanifold this is the projective space $\mb{CP}^{1|1}$, which is defined by taking $\mb{C}^{2|1}$ minus the $0|1$-dimensional locus where both even coordinates vanish, and quotienting by the equivalence relation $(u,v|\theta) \sim (\lambda u , \lambda v | \lambda \theta)$ where $\lambda$ is any non-zero complex  number \cite{Manin1991}. Taking $u,v|\theta$ as coordinates on $\mb{C}^{2|1}$, the superconformal coordinate charts $\bs z$ and $\bs w$ of \eq{ztoy} can be defined by
\begin{align}
z | \zeta & \equiv \frac{u}{v} \, \big| \frac{\theta}{v} & \text{ for }v & \neq 0 \, ; &
w | \psi & \equiv - \frac{v}{u} \, \Big| \frac{\theta}{u} & \text{ for }u & \neq 0 \, . \label{affinecharts}
\end{align}
It can be shown (see \eg~section 5.1.1 of \cite{Witten:2012ga}) that this is the only SRS \srs~whose reduced surface is the Riemann sphere.
\subsubsection{Super projective transformations}
\label{osp}
The superconformal structure on $\mb{CP}^{1|1}$ can also be defined by introducing a skew-symmetric bilinear form $
\langle  \cdot , \cdot \rangle $ on $\mb{CP}^{2|1}$ defined by (Eq.~(5.4) of \cite{Witten:2012ga})
\begin{align}
\langle Y , Y ' \rangle & = u v' - v u ' - \theta \theta ' \, , \label{bilin}
\end{align}
where $Y =( u,v|\theta)$ and $Y' = (u',v'| \theta')$. It can be shown (see, for example, Theorem (1.12) of Chapter 2 of \cite{Manin1991}) that the set of automorphisms of $\mb{CP}^{1|1}$ as a SRS is the supergroup which preserves the bilinear form \eq{bilin}, namely \osp. \osp~acts linearly on the homogeneous coordinates by $\text{GL}(2|1)$ matrices of
the form
\begin{align}
  \bs{\gamma}\, = \, \left( \begin{array}{cc|c} a & b & \alpha \\ c & d & \beta  \\
  \hline \gamma & \delta & e \end{array} \right) \, ,
\label{GLmatrix}
\end{align}
where the five even and four odd variables are subject to one even and two odd constraints as well as one even normalization condition:
\begin{align}
  \left( \begin{array}{c} \alpha \\ \beta \end{array} \right) & =
  \left( \begin{array}{cc} a & b \\ c & d \end{array} \right)
  \left( \begin{array}{c} - \delta \\ \gamma \end{array} \right) \, ,  &
  a d - b c - \alpha \beta & =  1 \, , &  e & = 1 - \alpha \beta \, ,
\label{OSpConstraints}
\end{align}
 so that \osp~has dimension $3|2$. The inverse of such a map $\bs \gamma$ is
  \begin{align}
  \bs \gamma^{-1} & =\left( \begin{array}{cc|c} d & - b & \delta \\ - c & a & - \gamma  \\
  \hline - \beta & \alpha  & e \end{array} \right) \, .
\label{ospinv}
  \end{align}
  In terms of the superconformal coordinates $\bs z \equiv z|\zeta$ defined in \eq{affinecharts}, $\bs{\gamma}$ acts as
 \begin{align}
 z | \zeta  \mapsto \bs{\gamma}(z|\zeta) & = \frac{az + b + \alpha \zeta}{c z + d + \beta \zeta} \, \Big| \, \frac{\gamma z + \delta + e \zeta}{c z + d + \beta \zeta} \, . \label{superproj}
 \end{align}
 The reduced form of $\bs{\gamma}$ obtained by setting $\zeta = 0$ in the even part of \eq{superproj} is just the M\"obius map $\gamma$ in \eq{moebius}.
 It is useful to introduce a bra-ket notation for $\mb{C}^{2|1}$ with respect to the bilinear form in \eq{bilin}. Given $Y = u,v|\theta \in \mb{C}^{2|1}$, let us define\footnote{The definition of $\bra{Y}$ differs by a minus sign from the one used in \cite{Magnea:2013lna,Magnea:2015fsa}.}
 \begin{align}
 \bra{Y} & = (- v\,\, u\, | \, -  \theta) \, , &
  \ket{Y} & = (u\,\, v \, |\, \theta ) \tran \label{braket1}
  \end{align}
  so that
  \begin{align}
  \langle Y | Y' \rangle & \equiv \sum_{i=1}^3\bra{Y}_i  \ket{Y ' }_i = \langle Y, Y' \rangle \, .\label{braket2}
 \end{align}
 There is a super-projective version of the cross-ratio of four points. Given four super-points $\bs{z}_i \in \mb{CP}^{1|1}$, $i=1, \ldots, 4$, we can define
 \begin{align}
 \wh{\Psi} ( \bs{z}_1 , \bs{z}_2 , \bs{z}_3, \bs{z}_4) & = \frac{ \langle \bs{z}_1 | \bs{z}_2 \rangle \langle \bs{z}_3 | \bs{z}_4 \rangle}{\langle \bs{z}_1 | \bs{z}_4 \rangle \langle \bs{z}_2 | \bs{z}_4 \rangle} \, . \label{sucr}
 \end{align}
 where  $\ket{\bs z_i} \in \mb{C}^{2|1}$ are homogeneous coordinates for the $\bs{z}_i$. Then $\wh{\Psi}$ is unchanged if we transform all four of the $\bs{z}_i \equiv z_i | \zeta_i$ according to \eq{superproj}.

 If we introduce the Neveu-Schwarz difference of two superpoints \cite{Fairlie:1973jw}:
 \begin{align}
 \bs z_1 \dm \bs z_2 & \equiv z_1 - z_2 - \zeta_1 \zeta_2 \, , \label{sdiff}
 \end{align} then the cross-ratio may also be written
 \begin{align}
  \wh{\Psi} ( \bs{z}_1 , \bs{z}_2 , \bs{z}_3, \bs{z}_4) & = \frac{ \bs{z}_1 \dm \bs{z}_2  }{ \bs{z}_1 \dm \bs{z}_4} \cdot \frac{ \bs{z}_3 \dm \bs{z}_4 }{  \bs{z}_2 \dm \bs{z}_3 } \, . \label{sucrsd}
 \end{align}
 A novelty in the super-projective case is that there is also an odd pseudo-invariant of three super-points (defined up to a sign) \cite{Hornfeck:1987wt}:
 \begin{align}
 \Theta(\bs{z}_1 , \bs{z}_2 , \bs{z}_3) & = \pm \frac{\zeta_1 (\bs z_2 \dm \bs z_3) + \zeta_2 (\bs z_3 \dm \bs z_1) + \zeta_3(\bs z_1 \dm \bs z_2) + \zeta_1 \zeta_2 \zeta_3 }{\sqrt{(\bs z_1 \dm \bs z_2)(\bs z_2 \dm \bs z_3)(\bs z_3 \dm \bs z_1)}} \label{oddinv} \, .
 \end{align}

\subsubsection{Hyperbolic superprojective maps}
\label{hyp21}
All elements of a bosonic Schottky group $G$ are hyperbolic, \ie~each one is conjugate by some \psl~map to a dilatation $z \mapsto k_\alpha z$. To define super Schottky groups we need a corresponding notion of a hyperbolic superprojective map (see Chapter 2, Section 2.10 of \cite{Manin1991}).
Consider the map $\bs{P}_\ve \in \osp$ given by
   \begin{align}
   \bs{P}_\ve & = \left( \begin{array}{cc|c} \ve & 0 & 0 \\ 0 & \ve^{-1} & 0 \\ \hline 0 & 0 & 1 \end{array}\right) \, , \label{spdef}
   \end{align}
   for some even $\varepsilon$ with\footnote{Inequalities are to be understood as holding modulo odd variables.} $|\ve| < 1$. In superconformal coordinates, $\bs{P}_\ve$ acts as
   \begin{align}
   z | \zeta \equiv \bs{z}   \mapsto  \bs{P}_\ve(\bs{z}) & = \ve^2 z | \ve \zeta  \, .  \label{zPe}
   \end{align}
    $\bs{P}_\ve$ has two fixed superpoints corresponding to the eigenvectors $(1,0|0)\tran$ and $(0,1|0)\tran$; denote them symbolically by $\bs z = \bs \infty $ and $\bs z  = \bs 0$, respectively (the odd third eigenvector of $\bs{P}_\ve$ is excluded from $\mb{CP}^{1|1}$ by definition). The first one is repulsive and the second one is attractive.

   Any map $\bs \gamma \in \osp$ that can be written in the form
   \begin{align}
   \bs \gamma & = \bs \Gamma^{-1} \bs P_\ve \bs \Gamma \, ;
   &
   \bs\Gamma & \in \osp \label{hypdef} \, ,
    \end{align}
    will be called \emph{hyperbolic}; we can denote the supermanifold of these maps by  $\hyp \subseteq \osp$ \cite{Manin1991}.
    We will call $\ve$ the \emph{semimultiplier} of $\bs{\gamma}$.

    Any $\bs \gamma \in \hyp$ has one attractive and one repulsive fixed superpoint: if $\bs \gamma$ is written in the form \eq{hypdef} then these are $\bs u = \bs \Gamma^{-1}(\bs 0)$ and $\bs v = \bs \Gamma^{-1} (\bs \infty)$ respectively. Given two super-points $\bs{u}$, $\bs{v} \in \mb{CP}^{1|1}$ whose even parts are distinct modulo odd variables, we can find a \hyp~map $\bs{\gamma}$ which has them as its attractive and repulsive fixed points, respectively. Just let $\bs \Gamma = \bs{\Gamma_{uv}}$ be any \osp~map with
   \begin{align}
   \bs{\Gamma_{uv}}(\bs{u}) & = \bs 0\, ,  &
   \bs{\Gamma_{uv}}(\bs{v}) & = \bs \infty  \, ; \label{gamcon}
   \end{align}
   one such \osp~matrix can be written using the bra-ket notation of \eq{braket1} as \cite{Magnea:2015fsa}:
   \begin{align}
     {\bs \Gamma}_{\bs {uv}} & =  \frac{1}{\sqrt{\langle {\bs u} | {\bs v} \rangle }}
  \left( \begin{array}{cc|c}
  u_2  & - u_1 & \wh \theta \\ \phantom{\Big|} v_2 & - v_1 & \wh\phi \\ \hline
  \frac{\phantom{\big|} u_2 \wh\phi - v_2 \wh\theta}{\sqrt{\langle {\bs u} | {\bs v} \rangle }} &
  \frac{\phantom{\big|} v_1 \wh\theta - u_1\wh \phi}{\sqrt{\langle {\bs u} | {\bs v} \rangle }} & \sqrt{\langle {\bs u} | {\bs v} \rangle } - \frac{3}{2} \frac{\phantom{\big|} \wh\theta\wh \phi}{\sqrt{\langle {\bs u} | {\bs v} \rangle }}
  \end{array} \right) \, ,
\label{Gammauvdef}
   \end{align}
   where $\ket{\bs u} = (u_1,u_2|\wh \theta)\tran$ and $\ket{\bs v} = (v_1,v_2| \wh \phi)\tran$.
   Then
   \begin{align}
   \bs{\gamma} & =  \bs{\Gamma_{uv}}^{-1}\,  \bs{P}_\ve \,  \bs{\Gamma_{uv}}
   \end{align}
   is the desired map.
   $\bs \gamma$ can be written using the bra-ket notation introduced in \eq{braket1} as \cite{Magnea:2015fsa}:
   \begin{align}
  {\bs \gamma} & = \bs 1  \, + \, \frac{1}{ \langle \bs{v} | \bs{u} \rangle }
   \Big( \left( 1 - \ve \right) \ket{\bs{v}} \bra{\bs{u}}
   \, - \, \left( 1 - \ve^{-1} \right) \ket{\bs{u}}
   \bra{\bs{v}} \, \Big) \, .
 \label{eq:SuSbraket}
   \end{align}
 \subsubsection{Super-Schottky groups}
 \label{ssg}
Essentially, the idea is to repeat the construction of section \ref{schot} but with the roles of the Riemann sphere and hyperbolic M\"obius maps being replaced, respectively, with the supersphere $\mb{CP}^{1|1}$ and the hyperbolic super-projective transformations described in section \ref{hyp21}.

So to build a SRS $\srs_g$ of genus $g$ (meaning one whose reduced surface $\surf_g$ is a RS of genus $g$), we give a group $\bs G \subseteq \hyp$ which is a free group on $g$ generators
   \begin{align}
   \{ \bs{\gamma}_i ,\,\, i = 1 , \ldots , g \} \, , &  &  \bs{\gamma}_i & \equiv \bs{\Gamma}_{\bs{u}_i\bs{v}_i}^{-1} \circ \bs{P}_{\ve_i }\circ  \bs{\Gamma}_{\bs{u}_i\bs{v}_i} \, .
   \end{align}
   $\bs{G}$ is a super Schottky group if the reduced group $\bs{G}\re \subseteq \psl $, obtained by setting all odd parameters of $\bs{G}$ to 0, is a Schottky group. The super Schottky covering space is $\bs{\Omega}(\bs{G}) = \pi^{-1} (\mb{CP}^1 - \Lambda(\bs{G}\re))$,  where $\pi : \mb{CP}^{1|1} \to \mb{CP}^{1}$ is the projection onto the even part.

   Each of the $g$ generators of $\bs{G}$ needs $3|2$ parameters to be specified, $\bs u_i = u_i | \theta_i$, $\bs v_i = v_i |\phi_i$, and $\ve_i$. But we can always make an \osp~change of coordinates so that $3|2$ of these are fixed to, say,
   \begin{align}
   \bs{u}_1 & = \bs{0} \, , &
   \bs{v}_1 & = \bs{\infty} \, , &
   \bs{v}_2 & = 1 | \phi_2 = 1 | \Theta(\bs{u}_1,\bs{v}_1 ,\bs{v}_2) \, . \label{canonsc}
   \end{align}
A marked super Schottky group whose fixed points satisfy \eq{canonsc} is said to be \emph{normalized}.   Let us denote the supermanifold of marked, normalized super Schottky groups of genus $g$ by  $\sscsp$; let us denote the canonical super Schottky moduli (coordinates for \sscsp) by
    \begin{align}
    \{\bs m^A\} & = \{ \ve_1, \ldots, \ve_g, u_2, \ldots , u_g ,v_3, \ldots, v_g| \theta_2 , \ldots, \theta_g, \phi_2, \ldots, \phi_g\} \, . \label{ssmod}
   \end{align}
   The dimension of \sscsp~is $3g-3|2g-2$.
   Of course, this is also the dimension of the supermoduli space of genus $g$ SRS, and in fact all SRS with even $\vartheta$ characteristics can be obtained in this way  \cite{Manin1991}.

It can be seen (Appendix \ref{SewingApp}) that the construction described here may be arrived at by starting with a supersphere and repeatedly gluing pairs of marked superpoints (\ie~Neveu-Schwarz punctures) which become the fixed superpoints, with the gluing parameters becoming the semimultipliers $\ve_i$.

   We will need the notion of a \emph{split} SRS. This means that the even coordinates $z_i$ are independent of the odd coordinates $\zeta_j$ in the transition functions between any overlapping superconformal charts $\bs z_i$, $\bs z_j$, \ie, that all transition functions are of the form $\bs z_i = f_{ij} ( z_j ) \big| \phi_{ij}( z_j | \zeta_j)$. In fact (see, for example, Section 2.1.2 of \cite{Witten:2012ga}), any split SRS $\srs_g$ is the total space of a line bundle $(\mathit{\Pi} K^{-1/2}) \to \surf_g$, where $\surf_g$ is the reduced surface of $\srs_g$, $ K^{-1/2}$ is the dual to some line bundle $K^{1/2}$ whose square is the canonical bundle $K$ of $\surf_g$, and the symbol $\mathit{\Pi}$ indicates that the fibres are taken to be Grassmann-odd. The choice of the square root $K^{1/2}$ of $K$ is called a \emph{spin-structure}.

   For an SRS given by a super Schottky group $\bs G$, the elements of $\bs G$ can be regarded as the transition functions of a suitably chosen atlas. If we write a generic element of $\bs G$ in the form \eq{superproj} then we see that the requirement of splitness imposes the condition $\alpha = \beta = 0 $ and hence $\gamma = \delta = 0 $ by \eq{OSpConstraints}. So for $\srs_g$ split, every element of $\bs G$ must be of the form
\begin{align}
\bs \gamma & = \left(\begin{array}{cc|c} a & b & 0 \\ c & d & 0 \\ \hline 0 & 0 & 1 \end{array}\right) \, ; & ad-bc & = 1 \, . \label{sl2}
\end{align}
The group of such matrices is \slc, not \psl~as in the bosonic case, because flipping the signs of $a,b,c,d$ in \eq{sl2} produces a distinct \osp~map. In particular, for $\bs \gamma \in \slc \cap \hyp$, this sign choice is equivalent to a sign choice for the semimultiplier $\ve$.

A marked super Schottky group $\bs G$ describes a split SRS $\srs_g$ if the odd components of the fixed superpoints which parametrize $\bs G$ vanish: $\bs u_i = u_i | 0$, $\bs v_i = v_i | 0 $. In this case, $\srs_g$ is isomorphic to the total space of the line bundle $\mathit{\Pi}\! K^{-1/2} \to \surf_g$, where $\surf_g$ is the RS given by the marked Schottky group $G$ parametrized by the fixed points $u_i$, $v_i$ and by the multipliers $k_i = \ve_i^2$. Since the RS $\surf_g$ depends only on $\ve_i^2$ not $\ve_i$, it is unaltered by changing any of the semimultipliers by a factor of $(-1)$. The same is not true for the line bundle $\mathit{\Pi}\! K^{-1/2}$: the sign choice for the semimultipliers $\ve_i$ parametrizing $\bs G$ fixes the spin structure, \ie~the choice of $K^{1/2}$. Recall that the spin structure of a marked Riemann surface of genus $g$ can be expressed in terms of a $\vartheta$ characteristic $(\vec{\epsilon}_a, \vec{\epsilon}_b) \in (\frac{1}{2} \mb{Z}/\mb{Z})^{2 g}$ whose parity is said to be even or odd depending on whether $4 \vec{\epsilon}_a \cdot \vec{\epsilon}_b$ is even or odd \cite{AlvarezGaume:1986es}. The signs of the semimultipliers $\ve_i$ fix $\vec{\epsilon}_B$ which describes the twisting of $K^{1/2}$ around the $B_i$ cycles of $\srs_g$. On the other hand, $\vec{\epsilon}_A$ is always zero with split super Schottky groups, so the spin structure's parity is always even.


  The abelian superdifferentials $\wh{\sigma}_i({\bs z}) = \wh{\phi}_i({\bs z}) [ \d {\bs z}]$ of a split SRS can be written down in terms of the abelian differentials $\omega_i(z) \d z$ of its reduced surface (normalized according to \eq{abelnorm}) as:
  \begin{align}
  \wh{\phi}_i(z|\zeta) & = \zeta \, \omega_i(z) \, .
  \end{align}
  The period matrix ${\bs \tau}_{ij}$ of a split SRS is equal to the period matrix $\tau_{ij}$ of its reduced surface (defined in \eq{permatdef}).

\section{Deformations with Schottky groups}
\label{defschot}
\subsection{Deformations with bosonic Schottky groups}
\label{bosdef}
Given a Riemann surface of genus $g$, described by a Schottky group, $\surf_g = (\rs - \Lambda(G))/G$, we will consider two different ways to describe a deformation of the complex structure to get a different surface $\surf_g'$ of the same genus.

The first way is simply to shift some of the Schottky moduli,
\begin{align}
m^a & \mapsto m^a + \delta m^a \, , & a & = 1 , \ldots, 3g - 3 \, ,  \label{moddef}
\end{align}
giving a new Schottky group $G'$ which gives $\surf_g'$ as the quotient $\surf_g' = (\rs - \Lambda(G'))/G'$, where the complex structure is inherited from the canonical one on $\mb{CP}^1$ in the usual way.

The second approach is to hold the Schottky group $G$ fixed, but to deform the complex structure away from the canonical one induced from $\rs$. This can be achieved by switching on a \emph{Beltrami differential} $\mu_{\tz}{}^{\, z} (\tz, z) \, \d \tz \otimes \partial_z$. We will discuss the relationship between the two approaches and see how we should interpret expressions involving Beltrami differentials in terms of the Schottky moduli.

Recall that on a smooth $d$-dimensional manifold, an almost complex structure is a tensor field $J_\mu{}^\nu$ satisfying $J_\mu{}^\rho J_\rho{}^\nu = - \delta_{\mu}^{\nu}$, and in $d=2$, this is always integrable to a complex structure. A function $w$ on the manifold is said to be holomorphic with respect to the complex structure if
\begin{align}
(J_\mu{}^\rho \partial_\rho - \ii \, \partial_\mu) w & = 0 \, . \label{holomdef}
\end{align}
 Now focusing on the $d=2$ case, we can always get a complex structure if we are given a Riemannian metric tensor $\d s^ 2  = g_{\mu \nu}\, \d \xi^\mu \otimes  \d \xi^\nu$: we set $J_{\mu}{}^\rho = \sqrt{\det g}\,  \epsilon_{\mu \sigma} g^{\sigma \rho}$ where $\epsilon_{\mu \sigma}$ is the antisymmetric symbol with $\epsilon_{12} = 1$ \cite{D'Hoker:1988ta}. If we have some local complex coordinates $(z, \tz)$ (which are \emph{not} necessarily compatible with the complex structure), we can write a general metric in the form $\d s^2 = \ex{2 \rho} | \d z + \mu_{\tz}{}^{\, z} (\tz, z) \d \tz |^2$. The almost complex structure which this metric induces can be inserted in \eq{holomdef} to give the holomorphicity condition for $w$ in the form
 \begin{align}
 \partial_\tz w & = \mu_\tz{}^{\, z} \partial_z w \, , \label{beleq}
 \end{align}
 which is called \emph{Beltrami's equation}. $\mu_\tz{}^{\, z}$ is called a Beltrami parameter or Beltrami differential. A complex function satisfying \eq{beleq} is called a uniformizing coordinate. Of course, if $\mu_\tz{}^{\, z} = 0$ then \eq{beleq} becomes the usual holomorphicity condition with respect to the local complex coordinates, $\partial_\tz w=0 $.

Let us define the coordinate deformation $\delta z (\tz , z)$ as the difference between a uniformizing coordinate $w(\tz, z)$ solving \eq{beleq} and the coordinate $z$:
\begin{align}
\delta z (\tz,z) & \equiv w(\tz, z) - z \, , \label{defdef}
\end{align}
then \eq{beleq} becomes
\begin{align}
\mu_\tz{}^{\, z}  & = \frac{\partial_{\tz} \delta z}{1 + \partial_z \delta z} \, . \label{mudelta}
\end{align}
To apply this to a Riemann surface described by a Schottky group $G$, we consider a Beltrami differential $\mu_\tz{}^{\, z}$ defined on $\rs$. In order for it to make sense on $\surf_g$ we require that $\mu_\tz{}^z$ transforms under $G$ as
\begin{align}
\mu_\tz{}^{\, z} \circ \gamma_\alpha & = \frac{\gamma_\alpha'}{(\gamma_\alpha')^*} \times \, \mu_\tz{}^{\, z} \, ,
\end{align}
for any $\gamma_\alpha \in G$, as well as that $\mu_\tz{}^{\, z}$ vanishes on the limit set $\Lambda(G)$. Then there is a unique function $w: \rs \to \rs$ satisfying \eq{beleq} and fixing $z=0,1,\infty$ ($w$ is then a \emph{normalized} uniformizing coordinate for $\mu_\tz{}^{\, z}$). Then the group
\begin{align}
G' & = w \, G \, w^{-1} \label{Gprime}
\end{align}
is also a Schottky group, and its limit set is $\Lambda(G') = w (\Lambda(G))$ \cite{Bers1975332}. Clearly, a choice of generators on $\{\gamma_i, \, i=1,\ldots,g\}$ for $G$ induces a set of generators $\{ \wt \gamma_i\}$ for $G'$, and thus we can find the associated shifts $\delta m^a$ of the Schottky moduli \eq{moddef}, since each modulus $m_a$ is defined as either a multiplier $k_i$ or a fixed point $u_i, v_i$ for one of the $g$ generators.

The converse also holds: given two sets of generators $\{\gamma_i\}$, $\{\gamma_i'\}$ defining marked Schottky groups $G_1$ and $G_2$ of the same rank, there exists a Beltrami parameter $\mu_\tz{}^{\, z}$ on $\rs$ for which the above statements are valid (see, for example, Proposition 1 of \cite{Bers1975332}).

Now, consider a small deformation in the moduli \eq{moddef} of a Schottky group $G$. There is an associated Beltrami differential and hence a unique normalized uniformizing coordinate $w$; let us consider the associated coordinate deformation $\delta z$ (\eq{defdef}). Let us define $\cc_a(\tz,z)$ by the following expansion of $\delta z$ for small values of $\delta m^a$:
\begin{align}
\delta z(\tz,z) & = \delta m^a\, \cc_a{}^z(\tz,z) + {\cal O}(\delta m^a)^2 \, . \label{cadef}
\end{align}
Then it follows from \eq{mudelta} that we have (to leading order) \cite{Roland:1993pm}
\begin{align}
\mu_\tz{}^{\, z} & = \delta m^a \, \partial_\tz \, \cc_a{}^z + {\cal O}(\delta m^a)^2 \, , \label{capot}
\end{align}
so that $\delta m^a  \cc_a{}^z(\tz,z)$ is (to leading order) a \emph{potential} for $\mu_\tz{}^{\, z}$, \ie~a function $F$ satisfying $\mu_\tz{}^{\, z}  = \partial_\tz F$. Note that $\cc_a{}^z(\tz,z)$ is not single-valued (as a vector field) under the Schottky group $G$ so it is not well defined on $\surf_g$.
To see this, we note that for any Schottky group element $\gamma_\alpha$ in $G$, we have from \eq{Gprime}
\begin{align}
\wt \gamma_\alpha (w(z)) & = w(\gamma_\alpha(z))  \label{TwwT} \, ,
\end{align}
where $\wt \gamma_\alpha $ is the matching element of $G'$. Rewriting $w$ in terms of $\cc_a$ using \eq{defdef} and \eq{cadef}, and using the Taylor series for $\wt \gamma_\alpha  = \gamma_\alpha[m^a + \delta m^a]$, we can expand both sides of \eq{TwwT} to first order in $\delta m^a$ getting
\begin{align}
\gamma_\alpha + \delta m^a \Big(\frac{\partial \gamma_\alpha}{\partial m^a}  + \gamma_\alpha'\,\, \cc_a{}^z \Big) & = \gamma_\alpha + \delta m^a \, ( \cc_a{}^z\circ \gamma_\alpha )+ {\cal O}(\delta m^a)^2 \nonumber
\end{align}
so
\begin{align}
\frac{1}{\gamma_\alpha' }(\cc_a{}^z\circ \gamma_\alpha)  -  \cc_a{}^z  & =  \frac{1}{\gamma_\alpha' } \frac{\partial \gamma_\alpha}{\partial m^a} \,  \equiv X_a[\gamma_\alpha] \, . \label{cocyc}
\end{align}
The function on the right-hand side $X_a[\gamma_\alpha](z)$ is always a quadratic polynomial, and the map $X_a$ is in fact a cocycle of $G$ representing an Eichler cohomology class $(X_a) \in H^1(G,\Pi[z]_{2})$.

The (first) Eichler cohomology group may be described as follows: we can define a right action of $G$ on $\Pi[z]_{2n-2}$, the vector space of polynomials $p(z)$ of degree $\leq 2n-2$, by
\begin{align}
p \mapsto p \cdot \gamma_\alpha \equiv  p(\gamma_\alpha(z)) \gamma_\alpha ' (z)^{1-n} \, ,
 \end{align}
 for a Schottky group element $\gamma_\alpha \in G$ and a polynomial $p \in \Pi[z]_{2n-2}$. The space of polynomials $\Pi[z]_{2n-2}$ is closed under this action because of the special property of M\"obius maps $\gamma$ that $\gamma(z)^{2n} \gamma'(z)^{-m}$ is a polynomial of degree $2m$ whenever $m \geq n$. Then a map $X:G \to \Pi[z]_{2n-2}$ is a 1-cocycle if
 \begin{align}
 X(\gamma_\alpha \gamma_\beta) & = X(\gamma_\alpha) \cdot \gamma_\beta + X(\gamma_\alpha)
 \shortintertext{and a 1-coboundary if}
  X(\gamma_\alpha)& = p \cdot \gamma_\alpha - p
   \end{align}
   for some polynomial $p \in \Pi[z]_{2n-2}$. The first cohomology group $H^1(G, \Pi[z]_{2n-2})$ is defined as the quotient space of 1-cocycles $Z^1(G, \Pi[z]_{2n-2})$ by 1-coboundaries $B^1(G, \Pi[z]_{2n-2})$ \cite{Gardiner74}.

The left-hand side of \eq{cocyc} satisfies the 1-cocycle condition for $n=2$. Furthermore, if the left-hand side of \eq{cocyc} were a 1-coboundary then $\cc_a{}^z - p$ would be single-valued as a vector field on $\surf_g$ for some polynomial $p$, and thus describe an infinitesimal change of coordinates, not a deformation of the complex structure. But since $\partial_\tz p(z) = 0$, $\cc_a{}^z - p$ would give the same Beltrami parameter as $\cc_a{}^z$ would (using \eq{capot}), so $\cc_a{}^z$ too would describe a trivial deformation. So we are really interested not in cocycles but rather elements of the cohomology group $H^1(G, \Pi[z]_{2})$. The dimension of this vector space is $3g-3$ \cite{Kra84}, so it makes sense to parametrize moduli deformations in this way.

A natural basis is given by computing the Eichler periods \eq{cocyc} of the $g$ Schottky generators $\gamma_\alpha = \gamma_i$ with respect to the $3g-3$ Schottky moduli. This gives \cite{Roland:1993pm}
\begin{align}
X_{k_i}[\gamma_j](z) & \equiv \frac{1}{\gamma_j'(z)} \frac{\partial \gamma_j}{\partial k_i} \Big|_{z} =  \frac{1}{k_i} \frac{(z-u_i)(z-v_i)}{u_i - v_i} \, \delta_{ij} \, , \nonumber \\
X_{v_i}[\gamma_j](z) & \equiv \frac{1}{\gamma_j'(z)} \frac{\partial \gamma_j}{\partial v_i} \Big|_{z} = -(1- k_i) \frac{(z-u_i)^2}{(v_i-u_i)^2} \, \delta_{ij} \, , \label{bosper} \\
X_{u_i}[\gamma_j](z) & \equiv \frac{1}{\gamma_j'(z)} \frac{\partial \gamma_j}{\partial u_i} \Big|_{z} = \frac{1-k_i}{k_i} \frac{(z - v_i)^2}{(v_i - u_i)^2} \,  \delta_{ij} \, ; \nonumber
\end{align}
$3g-3$ of these expressions are non-zero.

With the use of the periods \eq{bosper} we can simplify some surface integrals over $\surf_g$ involving Beltrami parameters in their integrands. If $f(z)$ is meromorphic on $\rs$ and holomorphic on $\rs - \Lambda(G)$, and transforms under $G$ as $f(T_\alpha(z)) = (T_\alpha'(z))^{-2}f(z)$ (\ie~so that $f(z) \, \d z \otimes \d z$ is a quadratic differential on $\surf_g$), then we may compute $\int_{\surf_g} \d^2 z \, \mu_\tz{}^z(\tz,z) f(z)$ in the following way. First we choose a fundamental region ${\cal F}$ for $G$, so the integral can be written with ${\cal F}$ as its domain. Next we use Stokes' theorem to replace the area integral with a contour integral over the boundary of the fundamental region, \ie~the Schottky circles ${\cal C}_i$, ${\cal C}_i'$ according to \eq{Fbound}, using \eq{capot} to rewrite $\mu_\tz{}^z$ in terms of $\delta m^a \cc_a{}^z$. With \eq{cocyc}, the contributions from pairs of Schottky circles ${\cal C}_i$, ${\cal C}_i'$ can be combined and expressed in terms of the periods \eq{bosper}. Lastly, since $f(z)$ and $X_a[\gamma_i](z)$ are holomorphic, the contour integral can be evaluated using Cauchy's integral formula. To leading order in $\delta m^a$, we have
\begin{align}
\int_{\Sigma_g} \d^2 z \, \mu_\tz{}^z(\tz, z) f(z) & = - \,
\delta m^a \int_{\cal F} \d \big( \cc_a{}^z(\tz, z) f(z) \, \d z \big) \, \nonumber  \\
& =  -  \,
\delta m^a \sum_{i=1}^g \Big[ \oint_{{\cal C}_i'} -  \oint_{{\cal C}_i} \Big]  \cc_a{}^z(\tz, z) f(z) \, \d z  \nonumber \\
& =  -  \,
\delta m^a \sum_{i=1}^g  \oint_{{\cal C}_i}  \bigg( \frac{ \cc_a{}^z(S_i(z)^*, S_i(z))}{S_i'(z)} -  \cc_a{}^z(\tz, z) \bigg) f(z) \, \d z \nonumber \\
& =  - \,
\delta m^a \sum_{i=1}^g  \oint_{{\cal C}_i}  X_a[\gamma_i](z) f(z) \, \d z \label{surfper}
\end{align}
using $\d^2 z \equiv  \d z \wedge \d \tz$.

This type of computation is useful in a number of ways: for example, Roland \cite{Roland:1993pm} computed the ghost zero mode contribution to the bosonic string integration measure on Schottky space, matching the result from sewing $N$-reggeon vertices, while McIntyre and Takhtajan used it in the construction of holomorphic functions on Schottky space from functional determinants \cite{McIntyre:2004xs} (some of which arise in string theory \cite{Martinec:1986bq}).
We will use \eq{surfper} to compute the variation of the period matrix with respect to the Schottky moduli.

If $\tau_{ij}$ is the period matrix of a marked Riemann surface $\surf_g$ with its canonical complex structure as defined in \eq{permatdef}, then by switching on a Beltrami differential $\mu_\tz{}^z$ on $\surf_g$ we get a second RS $\surf_g'$ whose period matrix is given by $\tau_{ij}'  = \tau_{ij} + \delta \tau_{ij} $ with
\begin{align}
 \delta \tau_{ij}  & = \frac{1}{(2 \pi \ii)^2} \int_{\surf_g} \d^2 z\, \omega_j(z)\, \mu_{\tz}{}^z(\tz,z)\, \omega_i(z) \, \, + \, \, {\cal O}(\mu_{\tz}{}^z)^2 \, , \label{rauch}
\end{align}
where $\omega_i(z) \d z$ are the abelian differentials \eq{abelnorm}. For a derivation of this formula, see, for example, section 2 of \cite{D'Hoker:2015fna}. Now, using \eq{surfper}, we arrive at the following expression for $\delta \tau_{ij}$:
\begin{align}
\delta \tau_{ij} & =\, - \, \delta m^a \,  \frac{1}{(2 \pi \ii )^2} \sum_{\ell=1}^g  \oint_{{\cal C}_\ell}  X_a[\gamma_\ell](z)\, \omega_j(z) \, \omega_i(z)\, \d z \, . \label{tauvarfor}
\end{align}
We can perform some checks of this formula. Let us focus on the genus $g=2$ case, and consider the variation of $\tau_{ij}$ with respect to $k_1$, the multiplier of $\gamma_1$. The associated Eichler period can be found by putting $u_1 = 0$ and $v_1 = \infty$ in the first line of \eq{bosper}, giving $X_{k_1}[\gamma_j](z) =  \delta_{1j}\, z / k_1$. Inserting this in \eq{tauvarfor} and using the expressions for the abelian differentials $\omega_i(z) \, \d z$ given in \eq{omg2}, we find that the variation of $\tau_{ij}$ is given by a contour integral around ${\cal C}_1$. With the abelian differentials expanded as power series in the multipliers $k_i$, the integrand's only pole inside ${\cal C}_1$ can be at $z=v_1 \equiv \infty$. In the limit as $\delta m^a = \delta k_1 \to 0$, we get
\begin{align}
\Big( \frac{\partial \tau_{ij}}{\partial k_1 }\Big) & =\, - \,  \frac{1}{2 \pi \ii} \left(  \frac{1}{2 \pi \ii} \oint_{{\cal C}_1}  \frac{z}{k_1} \, \omega_j(z) \, \omega_i(z)\, \d z \, \right) \nonumber \, ,
\\
& =  \frac{1}{2\pi \ii} \left( \begin{array}{cc} {1}/{k_1} &  - 2 \, k_2 \,{(1-u)^2(1-u^2)}/{u^2} \, \\
  - 2 \, k_2 \,{(1-u)^2(1-u^2)}/{u^2} & 2 {(1-u)^2}/{u}
 \end{array}\right)\,  \label{k1taudef}  \\
 & \hspace{250pt} + {\cal O}(k_1) + {\cal O}(k_2)^2 \, . \nonumber
\end{align}
Alternatively, \eq{k1taudef} could be computed directly by differentiating \eq{taug2}, which gives the same answer, so this is a check that our approach makes sense.

The variation of $(\tau_{ij})$ with  respect to the other two moduli, $k_2$ and $u$, may be similarly checked to give the expected result (in this case the relevant contour integrals would be around ${\cal C}_2$ because these moduli enter as parameters of $\gamma_2$, and the only pole of the integrand is at $z=v_2 \equiv 1$ after expanding in the multipliers $k_i$).
\subsection{Deformations with super Schottky groups}
\label{ssgdef}
In this section we apply the methods of the previous section to results of D'Hoker and Phong \cite{D'Hoker:1989ai} and Witten \cite{Witten:2012ga} to compute period matrices of non-split SRS as deformations from the split case.

We want to be able to describe deformations similarly to the approach in section \ref{bosdef}, where we used the concept of a Beltrami differential $\mu_\tz{}^z(\tz , z)$. It was crucial to be able to consider Beltrami differentials that were not holomorphic, exploiting the fact that on Riemann surfaces we could define natural anti-holomorphic coordinates $\tz$ as the complex conjugates of the holomorphic coordinates $\tz = z^*$, but on SRS we need to be more careful. A useful approach, which we will adopt here, is given in for example, section 3 of \cite{Witten:2012bh} and section 3.5.1 of \cite{Donagi:2014hza}. The idea is to proceed by embedding the SRS $\srs_g$ in a cs supermanifold\footnote{See \eg~section 4.8 of \cite{deligne} for the definition of cs supermanifolds, or \cite{Witten:2012bg} for an expository discussion.} $\wh\srs_g$ of dimension $2|1$ whose reduced space is the same Riemann surface $\surf_g$ as the reduced space of $\srs_g$ (as a smooth surface). For practical purposes, what this means is that $\wh\srs_g$ can be described by local coordinates $(\tz, z | \zeta)$ where, modulo any odd variables,  $\tz$ is the complex conjugate of $z$.

There is a notion of holomorphic functions on $\wh\srs_g$, namely, functions annihilated by
\begin{align}
\wt\partial & \equiv \d \tz \frac{\partial}{\partial \tz} \, , \label{srshol}
\end{align}
so any functions $f = f(z|\zeta)$ which depend only on $z$ and $\zeta$, and not on $\tz$, are holomorphic. Therefore holomorphic functions on $\wh\srs_g$ can be identified with holomorphic functions on the SRS $\srs_g$. \emph{Antiholomorphic} functions on $\wh\srs_g$ are locally functions of $\tz$, \ie~those annihilated by $\partial_z$ and $\partial_\zeta$; these vector fields generate a sub-bundle of the tangent bundle $T\wh\srs_g$ called the holomorphic tangent bundle, which can be naturally identified with the tangent bundle $T\srs_g$ of the SRS $\srs_g$.

Then in this framework, deformations of SRS are conceived of as deformations of the holomorphic structure of $\wh\srs_g$ which leave fixed the underlying cs supermanifold. This is analogous to how we can describe deformations of RS by leaving the underlying smooth manifold fixed but deforming the holomorphic structure with a Beltrami parameter $\mu_\tz{}^z$ according to \eq{beleq}.

To deform the holomorphic structure, we alter the holomorphicity condition $\wt \partial f = 0$ to $\wt \partial ' f = 0 $, where the new operator $\wt \partial ' $ is obtained by adding some $(0,1)$-form (\ie~a 1-form proportional to $\d \tz$) valued in $T\wh\srs_g$ to \eq{srshol}, in such a way that the underlying surface with the sheaf of functions which are holomorphic in this new sense is still an SRS. This means we need to consider deformations which do not alter the embedding of ${\cal D} \subseteq T \srs_g$; it is shown, for example, in section 3.5.3 of \cite{Witten:2012ga} that the general deformation with this property is
\begin{align}
\wt\partial  \mapsto \wt \partial\, ' & = \wt \partial \, +  \, \d \tz \, \Big( \,h_\tz{}^z(z, \tz ) \,\partial_z + \frac{1}{2} h_\tz{}^z(\tz, z)\zeta \partial_\zeta + \chi_\tz{}^\zeta(\tz,z) ( \partial_\zeta - \zeta \partial_z) \,  \Big) \, , \label{ew340}
\end{align}
where the perturbation is a $(0,1)$-form valued in ${\cal S}$, the sheaf of superconformal vector fields. The fields $h_\tz{}^z$ and $\chi_\tz{}^\zeta$ are usually known as a metric perturbation and a gravitino, respectively. We can combine them into a single superfield
\begin{align}
{\cal H}_\tz{}^z(\tz,z|\zeta) & = h_\tz{}^z(\tz,z ) \, + \,2 \,\zeta  \, \chi_\tz{}^\zeta(\tz,z) \, , \label{Hdef}
\end{align}
so then \eq{ew340} can be rewritten as
\begin{align}
\wt\partial  \mapsto \wt \partial\, ' & = \wt \partial \, +  \, \d \tz \, \Big(  {\cal H}_\tz{}^z \, \partial_z \, + \, \frac{1}{2} \big(D_\zeta {\cal H}_\tz{}^z \big) D_\zeta  \Big) \, , \label{sf340}
\end{align}
where $D_\zeta = \partial_\zeta + \zeta  \partial_z$ as usual. When $\srs_g$ is taken to be split, the superfield ${\cal H}_{\tz}{}^z$ can be separated into a metric perturbation and a gravitino globally, not only chart-by-chart (which would otherwise be the case).

We want to make contact between this and the super Schottky group description of deformations. From that point of view, deformations amount simply to shifting the parameters $\{ \bs m^A\}$ \eq{ssmod} of a normalized marked super Schottky group $\bs G$.
Let us consider a shift in the moduli $\bs m^A \mapsto \bs m^A + \delta \bs m^A$, defining a new normalized, marked super Schottky group $\bs G \mapsto \bs G'$. Let
\begin{align}
\bs w \equiv w|\psi: \wh{\mb{CP}}{}^{1|1} & \to \mb{CP}^{1|1} \, ,
&
\bs G ' & = \bs w \bs G \bs w^{-1} \label{supw}
\end{align}
be a map which preserves $\cal D$, but which is \emph{not} holomorphic with respect to the canonical holomorphic structure on its domain, which we consider as a cs supermanifold. $\bs w$ is a quasisuperconformal map \cite{Martinec:1986bq}, and is the SRS analogue of the map $w$ in \eq{Gprime}.

Let us define the coordinate deformation $ \delta \bs z$ using the difference of superpoints \eq{sdiff} by
\begin{align}
\delta \bs z(\tz,z|\zeta) &  \equiv \bs w (\tz,z|\zeta) \dm \bs z \nonumber
 \\
& =
 \delta  z(\tz,z|\zeta) + \zeta \, \delta \zeta(\tz,z|\zeta) \, , \label{defone}
\end{align}
where
\begin{align}
\delta z(\tz,z|\zeta) & = w(\tz,z|\zeta) - z \, ,
&
\delta \zeta(\tz,z|\zeta) & = \psi(\tz,z|\zeta) - \zeta \, . \label{deftwo}
\end{align}
Since $\bs w$ preserves ${\cal D}$, \eq{superconf} must hold (\ie~$D_\zeta w = \psi D_\zeta \psi$). To linear order in $\delta {\bs z}$, this is solved by \cite{verlindeHthesis}
\begin{align}
\delta \zeta & = \frac{1}{2} D_\zeta \delta \bs z \, + \,  {\cal O}(\delta \bs z^2) \, ; \label{deltazsuperconf}  &
\delta z & = \delta \bs z - \frac{1}{2} \zeta D_\zeta \delta {\bs z} \, + \,  {\cal O}(\delta \bs z^2) \, .
\end{align}
We posit that $\bs w$ is holomorphic with respect to some perturbed holomorphic structure on $\wh\srs_g$, so a holomorphic function $f$ on the deformed SRS $\srs_g'$ can be written locally as $f(\bs w) = f(w|\psi)$. Pulling $f$ back to $\wh\srs_g$, we get
 \begin{align}
 \wh{f}(\tz , z; \zeta) & \equiv f(\bs w(\tz,z|\zeta) )\, .  \label{fhat}
 \end{align}
 We want to restate the holomorphicity of $f$, \ie~the fact that it can be written only in terms of $\bs w$, in terms of the deformations \eq{defone} and \eq{deftwo} as functions on $\wh\srs_g$.
To first order in $\delta \bs z$, $\wh f$ can be Taylor expanded as
\begin{align}
\wh{f}(\tz , z; \zeta) & = f(\bs z) + \delta \bs z(\tz , z; \zeta) \, \partial_z \,f(\bs z) + \delta \zeta(\tz , z; \zeta) \, D_\zeta f (\bs z)  \, + \,  {\cal O}(\delta \bs z^2) \, . \label{fser}
\end{align}
This implies that at the leading order, $\partial_z \wh f = \partial_z f \, + \,  {\cal O}(\delta \bs z)$ and $\partial_\zeta \wh f = \partial_\zeta f \, + \,  {\cal O}(\delta \bs z)$. Using that fact, we act on \eq{fser} with $\wt\partial$ to obtain
\begin{align}
\d \tz \Big( \partial_\tz \,+\,(- \partial_\tz \delta \bs z )\, \partial_z\,+\,( - \partial_\tz \delta \zeta)\, D_\zeta \Big) \wh f & = 0 \, . \label{fhol}
\end{align}
Using \eq{deltazsuperconf}, we see that \eq{fhol} matches \eq{sf340} if the super Beltrami field ${\cal H}_{\tz}{}^z$ defined in \eq{Hdef} is related to $\delta {\bs z}$ by
\begin{align}
{\cal H}_{\tz}{}^z & = - \partial_\tz \delta \bs z  \, + \,  {\cal O}(\delta \bs z^2) \, .  \label{Hdz}
\end{align}
Thus, we can describe SRS deformations in terms of the superfield $\delta \bs z (\tz,z|\zeta)$ via \eq{Hdz}. It has been shown by Rabin that the description of deformations by $\delta \bs z$ is valid at first order when deforming a split SRS \cite{Rabin:1987pe}.
Although $ {\cal H}_\tz{}^z \d \tz \otimes \partial_z$ is well-defined on $\wh\srs_g$, \ie~it is single-valued under $\bs G$ as a vector-valued 1-form, the same is not true for $\delta \bs z$: it is only single-valued on the Schottky cover. In fact, just as in the bosonic case \eq{cocyc}, it is a cocycle of $\bs G$.

To see that $\delta \bs z$ is a cocycle, consider an arbitrary super Schottky group element $\bs \gamma_\alpha \in \bs G $, and let the corresponding element in the deformed group be $\wh{\bs \gamma}_\alpha \in \bs G' $. From \eq{supw} we have
\begin{align}
\bs w ({\bs \gamma}_\alpha (\tz,z|\zeta)) & =   \wh{\bs \gamma}_\alpha ( \bs w(\tz,z|\zeta)) \, . \label{supwT}
\end{align}
Let us write $\bs f \equiv \bs f^0 | \bs f^1 $ for the even- and odd-valued parts of a function.
Expanding both sides of \eq{supwT} to first order in $\delta \bs z$, we have
\begin{align}
   \bs \gamma_\alpha^0\,  +\, \delta z \circ \bs \gamma_\alpha & \,=\,\wh{\bs \gamma}_\alpha^0  \, + \,\delta\bs z \, \partial_z{\bs \gamma}_\alpha^0 \, + \,\delta\zeta  \,  D_\zeta {\bs \gamma}_\alpha^0   \, + \,  {\cal O}(\delta \bs z^2)\, , \label{defeven} \\
   \bs \gamma_\alpha^1\,  +\, \delta \zeta \circ \bs \gamma_\alpha & \,=\,\wh{\bs \gamma}_\alpha^1  \, + \,\delta\bs z \, \partial_z{\bs \gamma}_\alpha^1 \, + \,\delta\zeta  \,  D_\zeta {\bs \gamma}_\alpha^1   \, + \,  {\cal O}(\delta \bs z^2) \, . \label{defodd}
\end{align}
We can combine these to find the behaviour of $\delta \bs z$ under the super Schottky group: by definition \eq{defone} it transforms as
\begin{align}\delta \bs z \circ \bs \gamma_\alpha & \equiv  \delta z \circ \bs \gamma_\alpha +\bs \gamma_\alpha^1 \, \delta \zeta \circ \bs \gamma_\alpha \,  ,
\end{align}
then $\delta z \circ \bs \gamma_\alpha$ and $\delta \zeta \circ \bs \gamma_\alpha$ can be inserted from \eq{defeven} and \eq{defodd} yielding
\begin{align}
\delta \bs z \circ \bs \gamma_\alpha & = \wh{\bs \gamma}_\alpha^0 - {\bs \gamma}_\alpha^0 - \wh{\bs \gamma}_\alpha^1 \, {\bs \gamma}_\alpha^1  
+ \,\delta\bs z\, \partial_z{\bs \gamma}_\alpha^0 \, + \,\delta\zeta \, D_\zeta {\bs \gamma}_\alpha^0 \,
+ \, \bs \gamma_\alpha^1 \, \big( \delta\bs z\, \partial_z{\bs \gamma}_\alpha^1 \, + \,\delta\zeta \, D_\zeta {\bs \gamma}_\alpha^1 \big) \, .
\end{align}
The first three terms are just the difference of superpoints $\wh{\bs \gamma}_\alpha \dm \bs \gamma_\alpha$ (defined in \eq{sdiff}).
The coefficients of $\delta \zeta$ cancel because $D_\zeta {\bs \gamma}_\alpha^0 = {\bs \gamma}_\alpha^1 D_\zeta {\bs \gamma}_\alpha^1 $ since ${\bs \gamma}_\alpha$ is superconformal and thus satisfies \eq{superconf}. For the same reason, the coefficients of $\delta \bs z$ combine as the squared semijacobian of $\bs \gamma_\alpha$ since (using $\partial_z = D_\zeta^2 $),
\begin{align}
\partial_z \bs \gamma_\alpha^0 + \bs \gamma_\alpha^1 \partial_z\bs \gamma_\alpha^1 & = \big(D_\zeta {\bs \gamma}_\alpha^1 \big)^2 \, .
\end{align}
Thus we arrive at
\begin{align}
\frac{ \delta \bs z \circ \bs \gamma_\alpha }{( F_{{\bs \gamma}_\alpha} )^2 }  - \delta \bs z & = \frac{\wh {{\bs \gamma}_\alpha}  \dm \bs \gamma_\alpha}{( F_{{\bs \gamma}_\alpha} )^2 } \, + \, {\cal O}(\delta \bs z^2) \,  . \label{supper}
\end{align}
Now, if $\wh{\bs \gamma}_\alpha$ differs from $\bs \gamma_\alpha$ by deforming the matrix entries by two new odd parameter $\theta$, $\phi$, then the right hand side of \eq{supper} is a quadratic polynomial. That is to say, if we take $\bs \gamma_\alpha$ to be of the form of $\bs \gamma$ in \eq{GLmatrix}, then $\wh{\bs \gamma}_\alpha$ is obtained from it by the substitutions
\begin{align}
(a,b,c,d,e) & \mapsto ( a_+, b_+ , c_+ , d_+, e_+ ) =  (a,b,c,d,e) + {\cal O}(\theta \phi) \, , \\
(\alpha , \beta, \gamma, \delta)  & \mapsto (\alpha_+ , \beta_+,\gamma_+, \delta_+)  =  (\alpha , \beta, \gamma, \delta)  + {\cal O}(\theta) + {\cal O}(\phi) \,  .
\end{align}
subject to \eq{OSpConstraints}.
Then the first term on the right-hand-side of \eq{supper} is a polynomial in $z$ and $\zeta$ given by
\begin{align}
\frac{\wh {\bs \gamma}  \dm \bs \gamma}{( F_{\bs \gamma} )^2 } & = (  c_-\, z +  d_- \, + \beta_-\, \zeta )(a_+\, z + b_+ + \alpha_+ \, \zeta) - (c\, z+d+\beta\, \zeta)(a \,z + b + \alpha\, \zeta) \nonumber \\
& \hspace{150pt} - ( \gamma_- \, z +  \delta_-  + e \, \zeta)(\gamma \, z + \delta + e \,\zeta) \, , \label{quadrat}
\end{align}
where $x_-$ is defined by $x_- + x_+ \equiv 2x$ for $x=c,d,\beta,\gamma,\delta$.

A notion of Eichler cohomology (introduced after \eq{cocyc} above) makes sense for subgroups of \osp, in particular super Schottky groups. To define a group action, we can introduce a space of even-valued ``polynomials'' in a super-point $\bs z = z | \zeta$
\begin{align}
\Pi[\bs z]_{m} & \equiv  \{ p(\bs z)  = a_m z^m + \alpha_{m-1} z^{m-1} \zeta  + a_{m-1} z^{m-1} + \ldots + \alpha_0 \zeta + a_0 \} \, ,
\end{align}
where $\alpha_j$ and $a_j$ are odd and even coefficients.
Then it is easy to verify that there is a right-action of $\bs G$ on $\Pi[\bs z]_{2n-2}$ which takes the form
\begin{align}
 p \cdot \bs \gamma & \equiv \, F_{\bs \gamma}(\bs z)^{2-2n} \, p \circ \bs \gamma \, ,
\end{align}
where $F_{\bs \gamma}$ is the semijacobian of $\bs \gamma$ defined in \eq{superjac}.
A cocycle of $\bs G$ is a map $X: \bs G   \to \Pi[\bs z]_{2n-2}$ satisfying
\begin{align}
X (\bs \gamma_\alpha \bs \gamma_\beta) & = X(\bs \gamma_\alpha) \cdot \bs \gamma_\beta + X(\bs \gamma_\beta) \, ,
\end{align}
and a coboundary is a map $X: \bs G   \to \Pi[\bs z]_{2n-2}$ such that
for some $ p_X \in \Pi[\bs z]_{2n-2}$,
\begin{align}
X (\bs \gamma) & = p_X \cdot \bs \gamma - p_X \, ,
\end{align}
which is automatically a cocycle. Then the Eichler cohomology group $H^1( \bs G , \Pi[\bs z]_{2n-2})$ is the space of cocycles modulo the space of coboundaries.

We have seen in \eq{quadrat} that
\begin{align}
X_{\delta \bs z}(\bs \gamma_\alpha) & \equiv \frac{\wh {\bs \gamma}_\alpha  \dm \bs \gamma_\alpha}{( F_{\bs \gamma_\alpha} )^2 } \, \label{cocycdef}
\end{align}
 is always valued in $\Pi[\bs z]_2$ when the matrix entries of $\bs \gamma_\alpha$ and $\wh{\bs \gamma}_\alpha$ are equal modulo two odd constants $\theta$ and $ \phi$, and it is not hard to use the left-hand-side of \eq{supper} to check that the cocycle property is satisfied. Furthermore, similarly to the bosonic case, if the map were a coboundary then the associated super-Beltrami parameter would just describe a global change of coordinates, not a moduli deformation, so to describe deformations we're really interested in equivalence classes  $(X_{\delta \bs z}) \in H^1( \bs G , \Pi[\bs z]_2)$.

Now, let us consider a marked normalized super Schottky group $\bs G$ for a split SRS, so all of the elements are of the form \eq{sl2}. In particular this means that the odd parts of the fixed super-points $\bs u_i = u_i | \theta_i$ and $\bs v_i = v_i | \phi_i$ must be $\theta_i = \phi_i = 0$.
We want to `switch on' two odd parameters $\theta$ and $\phi$ and compute the Eichler periods \eq{cocycdef} associated to this deformation.

First of all, we consider the case where these two odd supermoduli are switched on via the two fixed superpoints of \emph{one} super Schottky group element, so the non-split super Schottky group $\bs G ' $ is generated by $(g-1)$ of the generators of $\bs G$ along with one non-split generator $\wh{\bs \gamma}_i$ which reduces modulo $\theta_i$, $\phi_i$ to $\bs{\gamma}_i$, the remaining split generator of $\bs G$. The non-split generator $\wh{\bs \gamma}_i$ has fixed points $\bs u_i= u_i|\theta_i$ and $\bs v_i  = v_i |\phi_i$ and supermultiplier $\ve_i$, with $\bs \gamma_i$ the same except $\theta_i = \phi_i =0 $. The Eichler cocyle $X_{\theta_i \phi_i}$ of the associated deformation $\bs G \to \bs G '$ is then computed from \eq{cocycdef}; it can be defined by giving the image of the $g$ generators $\bs \gamma_j$ of $\bs G$ as:
\begin{align}
X_{\theta_i \phi_i}[\bs \gamma_j] (\bs z)  & = \delta_{ij} \Big( (\ve_i-\ve_i^{-1}) \frac{(z-u_i)(z-v_i)}{(u_i-v_i)^2}  \theta_i \phi_i \label{percomp} \\
  & \hspace{140pt} + 2 (1-\ve_i^{-1})  \frac{(z-v_i)  \theta_i+\ve(z-u_i)  \phi_i}{u_i-v_i} \zeta \Big) \,  . \nonumber
\end{align}
The other possibility we will consider is that the two odd moduli enter via the fixed points of two \emph{different} super Schottky group elements. That is to say, the non-split super Schottky group $\bs G'$ is generated by $(g-2)$ of the generators of the split super Schottky group $\bs G$, along with two generators $\wh{\bs \gamma}_i$, $\wh{\bs \gamma}_j$, say, which each have one non-zero odd parameter --- for example, we could take $\theta_i \neq 0 \neq \theta_j$. In that case, the associated Eichler period $X_{\theta_i \theta_j}$ could be read off by adding two copies of \eq{percomp} with $\phi_i \to 0$ in each, yielding:
\begin{align}
X_{\theta_i \theta_j}[\bs \gamma_k](\bs z)  & = 2 \Big(  \delta_{ik} (1-\ve_i^{-1})  \frac{z-v_i }{u_i-v_i} \theta_i +  \delta_{jk} (1-\ve_j^{-1})  \frac{z-v_j }{u_j-v_j} \theta_j \Big) \zeta  \,  . \label{mixedint}
\end{align}
We could similarly use \eq{percomp} to write down expressions for the Eichler periods $X_{\phi_i \phi_j}$ and $X_{\theta_i \phi_j}$ where the two odd moduli are shared between two generators in different ways.

Now, just as we did in the bosonic case \eq{surfper}, we can use the Eichler periods to evaluate surface integrals involving the metric perturbation $h_{\tz}{}^z$ and the gravitino $\chi_{\tz}{}^\zeta$.

 Let us first suppose we have a surface integral over $\wh\srs_g$ whose integrand is a super-Beltrami coefficient ${\cal H}_\tz{}^z$ multiplied by a holomorphic function $f$ which transforms under the Schottky group as the coefficient of a $3/2$-superdifferential, \ie~with $f(\bs \gamma(\bs z)) = F_{\bs \gamma}(\bs z)^{-3} f(\bs z)$. Let us pick a fundamental domain in the Schottky covering space bounded by $2g$ circles, then with \eq{Hdz}, we can use the supermanifold version of Stokes' theorem (see \eg~section 3.4 of \cite{Witten:2012bg}) to rewrite the integral as a contour integral over the $2g$ Schottky circles. Using the transformation properties of $f$, the contributions from pairs of Schottky circles ${\cal C}_i$, ${\cal C}_i'$ can be grouped together and rewritten in terms of the Eichler periods we computed above, leaving us finally with a sum of $g$ meromorphic contour integrals (one for each $a_i$ cycle):
\begin{align}
\int_{\wh \srs_g} [\d \tz ; \d z | \d \zeta ] \, {\cal H}_\tz{}^z(\tz, z|\zeta) f(z|\zeta) & = - \, \int_{\cal F} \d \big( \delta \bs z (\tz, z | \zeta) f(z | \zeta) \, [\d z| \d \zeta] \big) \, \nonumber  \\
& =  -  \,
 \sum_{i=1}^g \Big[ \oint_{{\cal C}_i'} -  \oint_{{\cal C}_i} \Big]  \delta \bs z (\tz, z | \zeta) f(z | \zeta) \, [\d z| \d \zeta]   \nonumber \\
& =  -  \,
 \sum_{i=1}^g  \oint_{{\cal C}_i}  \bigg( \frac{ \delta \bs z \circ \bs \gamma_i }{(F_{\bs \gamma_i})^2} - \delta \bs z \bigg) f  \,  [\d z| \d \zeta]  \nonumber \\
& =  -  \,
 \sum_{i=1}^g  \oint_{{\cal C}_i}  X_{\delta \bs z}[\bs\gamma_i](\bs z) f(\bs z) \, [\d z | \d \zeta]  \, .  \label{surfper}
\end{align}
In the case that $\srs_g$ is split, we can repeat the computation in \eq{surfper}, isolating the two components in the $\zeta$-expansion of ${\cal H}_\tz{}^z$, \eq{Hdef}. Writing $f(z|\zeta) = f_0(z) + \zeta f_1(z)$, the Berezin integral can be carried out, yielding an integral on the reduced surface $\srs_{g,\text{red}}$ and the result is
\begin{align}
\int_{\srs_{g,\text{red}}} \!\!\!\!\!\!\d^2 z \, h_\tz{}^z(\tz, z) f_1(z) & =  -  \,
 \sum_{i=1}^g  \oint_{{\cal C}_i} X_h [\bs\gamma_i]( z) f_1(z) \, \d z \,  ,  \label{heich} \\
\int_{\srs_{g,\text{red}}}  \!\!\!\!\!\! \d^2 z \, \chi_\tz{}^\zeta (\tz, z)f_0(z)  & =  -  \,
 \sum_{i=1}^g  \oint_{{\cal C}_i}\, X_\chi [\bs\gamma_i]( z)  f_0(z) \,\d z  \,  , \label{chieich}
\end{align}
where the Eichler periods for the metric perturbation and the gravitino are polynomials in $z$ given by
\begin{align}
X_h & = X_{\delta \bs z}\Big|_{\zeta =0 } \, ,
&
 X_\chi  &= \frac{1}{2} \, \partial_\zeta  \,  X_{\delta \bs z} \, .
\end{align}
For example, when the non-splitness enters via a single super Schottky group generator as in \eq{percomp}, the Eichler periods are defined by their actions on the generators $\bs\gamma_j$ as:
\begin{align}
X_h[\bs \gamma_j](z)   & = \delta_{ij} \, (\ve_i-\ve_i^{-1}) \frac{(z-u_i)(z-v_i)}{(u_i-v_i)^2}  \theta_i \phi_i \, , \label{hper}
\shortintertext{and}
X_{\chi} [\bs \gamma_j] (z)  & =- \,  \delta_{ij} \, (1-\ve_i^{-1})  \frac{(z-v_i)  \theta_i+\ve(z-u_i)  \phi_i}{u_i-v_i} \,  . \label{chiper}
\end{align}
These can be used, for example, to compute the period matrix $\bs \tau_{ij} ' $ of a non-split SRS $\wh\srs_g'$ as the correction to the period matrix $\bs \tau_{ij}$ of a split SRS.

\subsubsection{The period matrix}
\label{permatsubs}
A procedure for writing down the period matrix of a non-split SRS in terms of a reduced surface with a gravitino field switched on is given, for example, by D'Hoker and Phong in section 6 of \cite{D'Hoker:1989ai}. We will follow the treatment by Witten in section 8 of \cite{Witten:2012ga}, although there the gauge choice $h_\tz{}^z =0 $ is used, which is not generically compatible with the parametrization of inequivalent superconformal structures by super Schottky groups. We have seen in \eq{hper} that switching on two odd super Schottky moduli $\theta_i$ and $\phi_i$ requires that $h_\tz{}^z$ has a non-zero Eichler period around the $B_i$ homology cycle, so certainly in that case we cannot assume that $h_\tz{}^z$ vanishes identically.

First of all, we need to construct a basis of $g$ holomorphic sections of the Berezinian bundle of $\wh \srs_g$. With respect to the holomorphic structure on $\wh\srs_g$ defined by $\partial_\tz$, the sections of this line bundle are given in superconformal coordinates by $\wh\phi_i (z|\zeta) [ \d z | \d \zeta]$, where $\phi_i(z|\zeta)$ is independent of $\tz$, and the 1-forms $\d z$ and $\d \zeta$ are a basis for $T^* \srs_g \subseteq T^* \wh \srs_g$. When the holomorphic structure is deformed, it is not only the coefficient functions $\wh \phi_i(z|\zeta)$ which have to be modified, but also the 1-forms $\d z$ and $\d \zeta$ which define $[\d z | \d \zeta]$.

To compute how the local basis $[ \d z | \d \zeta]$ of $\text{Ber}(\srs_g)$ transforms, it is useful to conceptualize it as a codimension-1 \emph{integral form} on $\wh\srs_g$. Integral forms were introduced in \cite{Leites77}; see section 3.2.3 of \cite{Witten:2012bg} for an introduction. To define an integral form on a supermanifold $M$ with local coordinate $t^i |\theta^j$, we consider the reversed-statistics tangent bundle $\PT M$. This is identical to the tangent bundle $TM$ with local coordinates given by the coordinates of $M$, $t^i| \theta^j$, as well as the 1-forms $\d t^i$ and $\d \theta^j$ as coordinates on the tangent spaces, except that $ \d t^i$ are taken to be anticommuting and $\d \theta^j$ are taken to be commuting. Differential forms are therefore just functions on $\PT M$ with polynomial dependence on the $\d \theta^i$'s. Integral forms on $M$ are distributions on $\PT M$ whose support is the locus $\d \theta^j = 0$.

In our case, $\PT \wh \srs_g$ is a $3|3$-dimensional supermanifold with local even coordinates $z,\tz,\d \zeta$ and odd coordinates $\d z, \d \tz, \zeta$. A differential form on $\wh \srs_g$ is a function on $\PT \wh \srs_g$ which is polynomial in $\d \zeta$, while an integral form on $\wh \srs_g$ is a distribution which vanishes for $\d \zeta \neq 0$. In particular, a section $\wh \sigma_i = \wh \phi_i(z|\zeta)[\d z | \d \zeta]$ of $\text{Ber}(\srs_g)$ can be associated with the integral form $\wh \phi_i(z|\zeta)\, \d z\, \delta( \d \zeta)$ on $\PT \wh \srs_g$.

To construct the space of holomorphic sections of the Berezinian bundle of the deformed surface $\srs_g' $, first we need to find what we should replace the symbol $[\d z | \d \zeta]$ with. The 1-forms $\d z$ and $\d \zeta$ are of type $(1,0)$, \ie~they have the property that their contraction with the vector field $\partial_\tz$ vanishes. Then in the presence of a deformed holomorphic structure \eq{ew340}, we need to find a new pair of 1-forms of type $(1,0)$. A computation shows that $\d z + (\chi_\tz{}^\zeta \, \zeta - h_\tz{}^z)\, \d \tz$ and $\d \zeta + (\chi_\tz{}^\zeta + \frac{1}{2} \zeta\, \partial_z h_\tz{}^z  )\, \d \tz $ have the required property, so a general section of $\text{Ber}(\srs_g')$ takes the form
\begin{align}
\wh{\sigma}_i & = \wh \phi_i(\tz,z|\zeta)
 \big[\, \d z + (\chi_\tz{}^\zeta \, \zeta - h_\tz{}^z)\, \d \tz \,\big|\, \d \zeta + (\chi_\tz{}^\zeta + \frac{1}{2} \zeta\, \partial_z h_\tz{}^z  )\, \d \tz \, \big] \label{pertber} \,
\end{align}
(for $h_\tz{}^z=0$ this is just Eq.~(8.19) of \cite{Witten:2012ga}).
Now, we want to find sections \eq{pertber} of $\ber(\srs_g')$ which are holomorphic. It is equivalent to require that the corresponding integral form on $\wh \srs_g$ is closed, or in other words that as a function on $\PT  \wh \srs_g$,
\begin{align}
\wh{\sigma}_i & = \wh \phi_i(\tz,z|\zeta)
 \big(\, \d z + (\chi_\tz{}^\zeta \, \zeta - h_\tz{}^z)\, \d \tz \,\big)\, \delta\big( \d \zeta + (\chi_\tz{}^\zeta + \frac{1}{2} \zeta\, \partial_z h_\tz{}^z  )\, \d \tz \, \big) \label{berPTM} \,
\end{align}
is annihilated by the odd vector field $\d$ defined by
\begin{align}
\d & \equiv \d \tz\, \partial_\tz + \d z \,\partial_z + \d \zeta \, \partial_\zeta \, . \label{ddef}
\end{align}
We can compute
\begin{align}
\d  \wh \sigma_i  & = - \d \tz\, \d z\, \delta ( \d \zeta)\, \Big( \partial_\tz\,\wh \phi_i - \partial_ z \big( \, \wh \phi_i \, (\chi_\tz{}^\zeta - h_\tz{}^z) \big) + \partial_\zeta \big( \, \wh \phi_i \, ( \chi_\tz{}^z + \frac{1}{2} \zeta \, \partial_z h_\tz{}^z ) \big) \Big) \, . \label{dsigma}
\end{align}
Expanding $\wh \phi_i$ in $\zeta$ as $\wh \phi_i(\tz,z| \zeta ) \equiv \wh \alpha_i(\tz,z) + \zeta \, \wh b_i(\tz,z) $, we can find with a computation that the vanishing of \eq{dsigma} is equivalent to the following pair of equations:
\begin{align}
\Big(\partial_\tz +h_\tz{}^z \partial_z    + \frac{1}{2} \big(\partial_z h_\tz{}^z  \big)\Big) \wh \alpha_i  + \wh b_i\, \chi_\tz{}^\zeta& = 0 \, ; \nonumber \\
\partial_\tz\, \wh b_i - \partial_z ( \wh \alpha_i\, \chi_\tz{}^\zeta - \wh b_i\, h_\tz{}^z ) & = 0 \, , \label{compeqs}
\end{align}
which are equivalent to the superfield equation
  \begin{align}
  \partial_\tz \wh \phi_i + D_\zeta \big( {\cal H}_\tz{}^z D_\zeta \wh \phi_i \,  +\, \frac{1}{2} (D_\zeta  {\cal H}_\tz{}^z )\, \wh \phi_i\big) & =0 \, .
  \end{align}
Note that~\eq{pertber}, \eq{dsigma} and \eq{compeqs} reduce to Eqs.~(8.19), (8.22) and (8.23) of Witten \cite{Witten:2012ga} for vanishing metric perturbation, $h_\tz{}^z = 0$. \eq{compeqs} is the same as Eq.~(3.3) of D'Hoker and Phong \cite{D'Hoker:2015kwa} with slightly different notation.

The idea is to solve \eq{compeqs} perturbatively in the odd variables which parametrize the deformation. The abelian superdifferential coefficients $\wh\phi_i(\tz,z|\zeta)$ on the deformed SRS $\srs_g'$ should be thought of as deformations of the the abelian superdifferential coefficients $\phi_i  (\tz,z|\zeta)$ on the split SRS $\srs_g$. For a split SRS, the coefficients of the abelian superdifferentials are locally of the form $\phi_i(\tz,z|\zeta) = \zeta \omega_i(\tz,z)$, where $\omega_i$ are the coefficients of abelian differentials on the reduced surface $\surf_g$ --- \ie, if we expand $\phi_i(\tz,z|\zeta)  = \alpha_i(\tz,z) + \zeta b_i(\tz,z)$ on the split SRS $\srs_g$, then we have $ \alpha_i =0 $ and $b_i = \omega_i$. We are considering a nilpotent deformation depending on two odd parameters $\theta$ and $\phi$, so we must have
\begin{align}
\wh \alpha_i & =  {\cal O}(\theta) + \cal{O}(\phi) \, ,
&
\wh b_i & =\omega_i + {\cal O}(\theta \phi) \, .
\end{align}
The PDE's \eq{compeqs} can be restated as integral equations over the reduced surface $\surf_g$ like so:
 \begin{align}
 \wh \alpha_i(\tz,z) & = - \frac{1}{2 \pi } \int_{\surf_g}  \d^2 w \, S(z,w) \Big[ \partial_w(\wh \alpha_i\, h_\tw{}^w) + \wh b_i\, \chi_\tw{}^\psi - \frac{1}{2} \wh \alpha_i\, \partial_w h_\tw{}^w \Big](\tw,w) \, ; \label{alpheq} \\
 \wh b_i (\tz,z) & = \omega_i(z) - \frac{1}{2 \pi } \int_{\surf_g} \d^2 w \, \partial_z \partial_w \log E(z,w) \, \big[ \wh \alpha_i\, \chi_\tw{}^w - \wh b_i \, h_\tw{}^w \big] (\tw,w) \, . \label{beq}
 \end{align}
 Here $S(z,w)$ is the Szeg\H o kernel $S(z,w)$, which is a meromorphic section of $K^{1/2}_z \otimes K^{1/2}_w$ characterized by
  \begin{align}
  \partial_\tz S(z,w) & = 2 \pi \,\delta^2(z,w) \, ,
  &
  S(z,w) & = - S(w,z) \, ,
  \end{align}
  and $E(z,w)$ is the Schottky-Klein prime form, which is a holomorphic $(-1/2,0)$ form in both $z$ and $w$ with a unique zero at $z=w$.
  Now, the idea is to solve \eq{alpheq} and \eq{beq} iteratively order-by-order in the odd parameters. But since we are restricting ourselves to two odd parameters $\theta$, $\phi$, this procedure terminates at the first step. We note that we have
  \begin{align}
  \wh \chi_\tz{}^\zeta & = {\cal O}(\theta) + {\cal O}(\phi) \,
  &
  \wh h_\tz{}^z & = {\cal O}(\theta \phi) \,  ,
  \end{align}
  so any terms containing both $\wh \alpha_i$ and $h_\tz{}^z$ necessarily vanish, and $\wh b_i$ may be replaced with $\omega_i$ anywhere it is multiplied by either $h_\tz{}^z$ or $\chi_\tz{}^\zeta$. So in this case, the integral equations reduce to
 \begin{align}
 \wh \alpha_i(\tz,z) & = - \frac{1}{2 \pi } \int_{\surf_g}  \d^2 w \, S(z,w) \,\omega_i(w)\, \chi_\tw{}^\psi(\tw,w) \, ; \label{alpheq2} \\
 \wh b_i (\tz,z) & = \omega_i(z) - \frac{1}{2 \pi } \int_{\surf_g} \d^2 w \, \partial_z \partial_w \log E(z,w) \, \big[ \wh \alpha_i\, \chi_\tw{}^\psi - \omega_i \, h_\tw{}^w \big] (\tw,w) \, . \label{beq2}
 \end{align}
 The right-hand side of \eq{alpheq2} is exactly of the form of the left-hand side of \eq{chieich}, so $\wh \alpha_i$ is given by the contour integral
 \begin{align}
 \wh \alpha_i(\tz,z) & =  \frac{1}{2 \pi \ii}  \,
 \sum_{i=1}^g  \oint_{{\cal C}_i} \, S(z,w) \,\omega_i(w) X_\chi [\bs\gamma_i](w)  \,\d w  \, . \label{alphcont}
 \end{align}
 Similarly, we can write down a formula for $\wh b_i$ in terms of \eq{heich} and \eq{chieich}, although we only need to know \eq{alphcont} in order to write down the period matrix.

 The formula for the period matrix can be written down following section 8.3 of ref.~\cite{Witten:2012ga}. We can find $g$ even closed 1-forms on $\surf_g$:
 \begin{align}
 \wh\rho_i \,&=\, \wh b_i \, \d z \, + \, (\wh \alpha_i \,\chi_\tz{}^\zeta-\wh b_i \, h_\tz{}^z )\, \d \tz \, , \label{rhodef}
 \end{align}
 which satisfy $\d \wh \rho_i = 0$ because of the second equation in \eq{compeqs}. Integrating \eq{rhodef} around the $B_j$ homology cycle and using Riemann's bilinear relation, we find that the period matrix $\bs \tau_{ij} '$ of the deformed SRS $\srs_g'$ is given in terms of $\bs\tau_{ij}$, the period matrix of the split surface $\srs_g$, by
 \begin{align}
 \bs \tau_{ij} '  & = \bs \tau_{ij}  \, - \, \frac{1}{(2 \pi \ii)}  \int_{\surf_g} \d^2 z \,  \omega_j\,(\wh \alpha_i \,\chi_\tz{}^\zeta-\wh b_i \, h_\tz{}^z )\, . \label{sutaudef}
 \end{align}
 With some manipulations, we can write this in terms of Eichler periods of $\bs G$. Specializing to the case with two odd supermoduli switched on, we find that the integrand in \eq{sutaudef} is a total derivative. On the Schottky cover, let us write $h_\tz{}^z = - \partial_\tz \delta z^0$ and $\chi_\tz{}^\zeta = - \partial_\tz \delta \zeta^0 $, where we can take $\delta z^0 \in {\cal O}(\theta \phi)$ and $\delta \zeta^0 \in {\cal O}(\theta) + {\cal O}(\phi)$. Then we have
 \begin{align}
 \omega_j\, \d z\wedge(\wh \alpha_i \,\chi_\tz{}^\zeta-\wh b_i \, h_\tz{}^z )\, \d \tz  \, & =
 \d \big( \omega_j\, \d z(\wh \alpha_i \,\delta \zeta^0 -\wh b_i \, \delta z^0 )\big) \, . \label{pmtotder}
 \end{align}
 To see why this holds, note that when the exterior derivative on the right-hand side is expanded, the term proportional to $\partial_\tz \wh b_i\, \delta z^0 $ vanishes because $\wh b_i$ is the sum of a holomorphic part $\omega_i$ and a nilpotent part. Similarly,  \eq{compeqs} implies that with only two odd supermoduli we have $\partial_\tz \wh \alpha_i =  \wh b_i \, \partial_\tz\delta \zeta^0$, so the term including $\partial_\tz \wh \alpha_i\, \delta \zeta^0 $  is proportional to $\partial_\tz \delta \zeta^0 \,  \delta \zeta^0 \, \propto\, \partial_\tz( \delta \zeta^0)^2 = 0$ and it vanishes too.

  From the single-valuedness of $\wh \rho_i$ in \eq{rhodef}, it follows that $\wh b_i$ and $\wh \alpha_i$ transform under $\bs G$ as superdifferentials of weight 1 and 1/2, respectively, \ie~for $\bs \gamma \in \bs G$ we have
 \begin{align}
 \wh b_i (\bs \gamma(\bs z)) & = { F_{\bs \gamma}(\bs z)^{-2}}\,\,  \wh b_i( \bs z) \, ,
 &
 \wh \alpha_i (\bs \gamma(\bs z)) & = { F_{\bs \gamma}(\bs z)^{-1}}\,\,  \wh \alpha_i( \bs z) \label{abpers} \, ,
 \end{align}
 so the only objects on the right-hand side of \eq{pmtotder} which are not single-valued on $\srs_g$ are $\delta z^0$ and $\delta \zeta^0$. Then, similarly to the calculation in the bosonic case \eq{tauvarfor}, Stokes' theorem in the form of \eq{heich} and \eq{chieich} can be used to write \eq{sutaudef} in terms of the Eichler periods for $h_\tz{}^z$ and  $\chi_\tz{}^\zeta$. We get
 \begin{align}
 \bs \tau_{ij} '  & = \bs \tau_{ij}  + \frac{1}{(2 \pi \ii)^2}\sum_{i=1}^g  \oint_{{\cal C}_i} \d z \, \omega_j   \Big(\wh \alpha_i\, X_\chi [\bs\gamma_i] -\wh b_i\,  X_\chi [\bs\gamma_i]\, \Big) \, . \label{taufromdef}
 \end{align}
 Now, we are considering a super-Beltrami coefficient ${\cal H}_\tz{}^z = h_{\tz}{}^z + 2 \,\zeta\, \chi_\tz{}^\zeta$ which describes a deformation of $\srs_g \to \srs_g'$ away from the split locus, which amounts to `switching on' two odd moduli (\ie, choosing non-zero values for the odd coordinates of the fixed super-points of $\bs G$).
 But the super-Schottky group formula for the period matrix \eq{schopm} is valid regardless of whether $\bs G$ is split. This gives us two different formulae for the same period matrix, and we can check that they match by computing the first few terms in the power-series expansion in the semimultipliers $\varepsilon_i$.

\subsubsection{Computing the period matrix in genus $g=2$}
\label{norm}
  First of all let us take a genus $g=2$ SRS $\srs_2'$ described by a normalized super Schottky group $\bs G'$ as a deformation of a split SRS $\srs_2$ described by a super Schottky group $\bs G$. Both $\bs G$ and $\bs G'$ have the same even moduli, \ie~the fixed point $u_2 \equiv u$ and the supermultipliers $\varepsilon_1$ and $\varepsilon_2$. But the two odd moduli $\theta_2 \equiv \theta$, $\phi_2 \equiv \phi$ are `switched on' for $\bs G '$ and set to zero for $\bs G$. The Eichler periods associated to this deformation are
  \begin{align}
X_h[\bs \gamma_i](z)   & = \delta_{i2} \, (\ve_2-\ve_2^{-1}) \frac{(z-u)(z-1)}{(1-u)^2} \, \theta \phi \, , \label{hper2}
\shortintertext{and}
X_{\chi} [\bs \gamma_i] (z)  & = \,  \delta_{i2} \, (1-\ve_2^{-1})  \frac{(z-1) \, \theta\,+\,\ve_2\,(z-u) \, \phi}{1-u} \,  . \label{chiper2}
\end{align}
So to find $\wh{\alpha}_i$, we can insert the Eichler period $X_{\chi}$ from \eq{chiper2} into the contour integral formula \eq{alphcont}. The other ingredients needed for the computation are the abelian differentials on the reduced surface which can be computed from \eq{schoab}
and are given to second order in $\varepsilon_i$ by putting $k_i \to \ve_i^2$ in \eq{omg2}, and the Szeg\H{o} kernel which is given by the formula in \eq{szego1}, with the first few terms in the $\ve_i$-expansion for genus $g=2$ written down in \eq{Szwg2}. We use
\begin{align}
 \wh \alpha_i(\tz,z) & = \frac{1-\ve_2^{-1}}{1-u}  \frac{1}{2 \pi \ii}   \oint_{{\cal C}_2}  \d w \, \,S(z,w) \,\,\omega_i(w) \, \,\big( (w-1) \, \theta\,+\,\ve_2\,(w-u) \, \phi \big)\, , \label{alphcont2}
 \end{align}
 and power-expand in the semimultipliers $\ve_i$, then the integrand has a unique simple pole at $v_2\equiv 1$. The resulting expression for $\wh \alpha_i$ matches the one computed directly from the Poincar\'e series \eq{absubk} given in \eq{absudnorm}.

With the expression for $\wh \alpha_i(z)$ to hand, the period matrix $(\bs \tau_{ij})$ can be written down from the formula \eq{taufromdef}. Both Eichler periods \eq{hper2} and \eq{chiper2} for this deformation contain a $\delta_{i2}$ so again the formula reduces to a single contour integral around ${\cal C}_2$, and because we have expanded in $\ve_i$ the only simple pole is at the fixed point $v_2 \equiv 1$. So we find
\begin{align}
\bs \tau_{ij} ' & = \bs \tau_{ij} + \frac{1}{(2 \pi \ii)^2} \oint_{z=1} \! \! \! \!\d z \, \, \omega_j(z) \, \Big( \wh \alpha_i(z) \, \big((z-1) \, \theta\,+\,\ve_2\,(z-u) \, \phi\big)\frac{1-\ve_2^{-1}}{1-u} \label{pmg2int} \\
& \hspace{180pt} + \, \omega_i(z) \, (z-u)(z-1) \frac{\ve_2-\ve_2^{-1}}{(1-u)^2} \, \theta \phi \Big)\nonumber \, .
\end{align}
This can be evaluated and matches the expression in \eq{pmnormg2} computed directly from the super Schottky group series \eq{schopm}.
\subsubsection{An alternative choice of odd supermoduli}
\label{shared}
The standard prescription \eq{canonsc} for the canonical normalization of a super Schottky group means that the $g$ generators $\bs \gamma_i$ are not all on the same footing, since $\bs \gamma_1$ has no odd parameters while the other generators have two. In the genus $g=2$ case, there are two odd supermoduli in total so a normalization in which both generators have one odd parameter would be more symmetric. Let us use \osp~invariance to normalize the odd parts of the fixed superpoints as
  \begin{align}
  \theta_1 & = \xi \, ;
  &
  \theta_2 & = \theta \, ;
  &
  \phi_1 = \phi_2 & = 0 \, . \label{sharedodd}
  \end{align}
  so now $\theta$ and $\xi$ are the odd supermoduli. The even supermoduli are the same as in the canonically normalized case. The Eichler periods for the deformation from the split case amounting to `switching on' these two odd supermoduli can be found from \eq{hper} and \eq{chiper}, and are given by
  \begin{align}
  X_\chi[ \bs \gamma_1 ] & = \big( {\ve_1^{-1}} - 1 \big) \, \xi \, , &
  X_\chi[ \bs \gamma_2 ] & = \big( {\ve_2^{-1}} - 1 \big) \frac{1-z}{1-u}\, \theta \, ,
  &
  X_h[ \bs \gamma_i ] & = 0 \, , \label{Eichshared}
  \end{align}
  so  we can take this deformation to be parametrized by the gravitino alone, with the metric perturbation set to zero as in Section 8.3 of \cite{Witten:2012ga}.

  As in Section \ref{norm}, the Eichler periods \eq{Eichshared} can be used to compute the abelian superdifferentials $\wh \phi_i(z|\zeta)$ and the period matrix. The results are the same as in the expressions in \eq{absudshar} and  \eq{pmsharedg2}, respectively, which are computed directly from the super Schottky group series formulae \eq{absubk} and \eq{schopm}.
\section*{Acknowledgements}
I thank R.~Russo and S.~Sciuto for useful discussions and contributions at an early stage of the work, and also L.~Magnea for collaboration on related projects. Helpful comments on the manuscript have been given by J.~Hayling, E.~Hughes, C.~Maccaferri and M.~Moskovic. This work was supported by the Compagnia di San Paolo contract
``MAST: Modern Applications of String Theory'' \verb=TO-Call3-2012-0088=.
\appendix
\section{Appendix: Selected (super) Schottky group formulae}
\label{schotapp}
 \subsection{Abelian differentials and the period matrix}
 \label{abdifpm}
On a compact Riemann surface $\surf_g$, there is a $g$-dimensional vector space of abelian differentials, \ie~1-forms which can be written locally as $\varphi(z) \d z$ where $\varphi$ is holomorphic. A normalized basis $\omega_i$, $i=1,\ldots,g$  can be fixed by specifying the periods around the $A^i$-cycles,
\begin{align}
\frac{1}{ 2 \pi \ii } \oint_{A^i} \omega_j(z) \, \d z & =   \delta_{ij} \, . \label{abelnorm}
\end{align}
Then the \emph{period matrix} $\tau_{ij}$ is a $g \times g$ matrix given by the integrals of $\omega_i$ around the $B_j$ cycles:
\begin{align}
\tau_{ij} & = \frac{1}{2 \pi \ii} \oint_{B_j} \omega_{i} (z) \, \d z \, . \label{permatdef}
\end{align}
The Riemann bilinear relations show that the period matrix $\tau_{ij}$ is symmetric $\tau_{ij} = \tau_{ji}$ and has positive-definite imaginary part, $\text{Im}(\tau_{ij}) > 0$ \cite{gunning}.

For a surface $\surf$ given by a Schottky group $G$, the abelian differentials can be expressed as Poincar\'e $\theta$-series: we may take
\begin{align}
\omega_i(z) \, \d z & = { \sum_{\gamma_\alpha }}^{(i)} \Big( \frac{1}{z - \gamma_\alpha(u_i)} - \frac{1}{z - \gamma_\alpha(v_i)} \Big) \, \d z \, , \label{schoab}
\end{align}
where the summation ${ \sum_{\gamma_\alpha }}^{(i)}$ is over all elements of the Schottky group $G$ which do not have $\gamma_i^{\pm n}$ as their right-most factor.
 Note that the summand on the right-hand side of \eq{schoab} can be written in terms of cross-ratios as
 \begin{align}
  \Big( \frac{1}{z - \gamma_\alpha(u_i)} - \frac{1}{z - \gamma_\alpha(v_i)} \Big) \, \d z & = \d \log \Psi(z,\gamma_\alpha(u),a,\gamma_\alpha(v)) \, , \label{abelcross}
  \end{align}
  where $a$ is an arbitrary point and we have used the notation
  \begin{align}
  \Psi(z_1,z_2,z_3,z_4)  & \equiv \frac{z_1 - z_2}{z_1 - z_4}\frac{z_3 - z_4}{z_2 - z_3} \, . \label{crdef}
 \end{align}
 Using \eq{abelcross}, we can compute the periods of the abelian differentials \eq{schoab} around the $B_i$ cycles as in \eq{permatdef}, giving the following Schottky group expression for $\tau_{ij}$:
\begin{align}
\tau_{ij} & = \frac{1}{2 \pi \ii} \Big( \delta_{ij} \log k_i \,  - \, {\vphantom{\sum_{\gamma_\alpha}}}^{(i)} {\sum_{\gamma_\alpha}}^{'(j)} \log \frac{u_j - \gamma_\alpha(v_i)}{u_j - \gamma_\alpha(u_i)}\frac{v_j - \gamma_\alpha(u_i)}{v_j - \gamma_\alpha(v_i)} \Big) \, ,  \label{boschopm}
\end{align}
where the summation ${\vphantom{\sum_{\gamma_\alpha}}}^{(i)} {\sum_{\gamma_\alpha}}^{'(j)}$ is over all elements of the Schottky group which have neither $\gamma_i^{\pm n}$ as their left-most factor nor  $\gamma_j^{\pm m}$ as their right-most factor, also excluding the identity when $i=j$. A detailed computation of the expression for $\tau_{ij}$ in \eq{boschopm} is given in section 7 of \cite{Mandelstam:1985ww}. This expression arises automatically in the approach to computing multi-loop string amplitudes by sewing $N$-reggeon vertices in the operator formalism \cite{DiVecchia:1988cy}.

\subsection{Prime form and Szeg\H o kernel}
\label{pfsk}
There are Schottky group formulae for the Schottky-Klein prime form and the Szeg\H o kernel, both defined after \eq{beq}. The prime form is given by \cite{DiVecchia:1988cy}
  \begin{align}
  E(z,w) & = \frac{z-w}{\sqrt{\d z} \sqrt{\d w}} {\prod_{\gamma_\alpha}}' \frac{z - \gamma_\alpha(w)}{z - \gamma_\alpha(z)}\frac{w - \gamma_\alpha(z)}{w - \gamma_\alpha(w)} \, , \label{Ezw}
  \end{align}
  where the product over the Schottky group excludes the identity, and only includes an element $\gamma_\alpha \in G$ if its inverse $\gamma_\alpha^{-1}$ is excluded (the choice between $\gamma_\alpha$ and $\gamma_\alpha^{-1}$ is arbitrary).
  In formulae such as \eq{beq} for the abelian differentials in the presence of a deformation, the prime form enters in the form $\partial_z \partial_w \log  E(z,w)$. This can be computed from \eq{Ezw} as
  \begin{align}
 \partial_z \partial_w \log  E(z,w) & = \sum_{\gamma_\alpha} \frac{ \gamma_\alpha ' (w)}{(z-\gamma_\alpha(w))^2} \, , \label{ddlnEzw}
  \end{align}
  where the sum is now over the whole Schottky group $G$.

  A Schottky group formula for the Szeg\H o kernel was given by Pezzella in
  \cite{Pezzella:1988jr}:
  \begin{align}
  S(z,w) & = \sum_{{\gamma_\alpha}} \frac{\gamma_\alpha'(w)^{1/2}}{z - \gamma_\alpha(w) } \,   \label{szego1}
  \end{align}
  (the terms in this series are the square roots of the terms in \eq{ddlnEzw}).
  The branch of the square root in \eq{szego1} is fixed by
  \begin{align}
  (\gamma_\alpha'(z))^{1/2} & \equiv \frac{1}{c_\alpha z + d_\alpha }\,  , &
  \gamma_\alpha & = \left(\begin{array}{cc} a_\alpha & b_\alpha \\ c_\alpha & d_\alpha \end{array}\right) \, , & a_\alpha d_\alpha - b_\alpha c_\alpha & = 1 \, .
  \end{align}
  In order for this specification to be unambiguous, it is necessary to consider the Schottky group $G$ as a subgroup not of $\psl$ but of its double cover $\slc$, so we may no longer multiply all entries of a matrix $\gamma_\alpha \in G $ by $(-1)$. We must make a choice of sign for the $g$ generators, then the other signs are fixed algebraically. This sign choice corresponds to a choice of $\vartheta$-characteristic $\vec \epsilon_b \in (\frac{1}{2} \mb{Z}/(2\mb{Z}))^g$ determining the spin structure around the $B_i$-cycles.

  If we embed \slc~in \osp~by writing the matrices in the form \eq{sl2}, then we can use super-Schottky group notation to rewrite \eq{szego1} in a way which makes the symmetry properties more transparent, as
  \begin{align}
  S(z,w)
  & =  \sum_{\gamma_\alpha} \frac{\bra{z} \Phi \, \gamma_\alpha \, \Phi \ket{w}}{\bra{z}\, \gamma_\alpha \, \ket{w}} \, ,  \label{szego2}
  \end{align}
  where in the last line we have used the bra-ket notation \eq{braket1} with $\bra{z}  = (-1,z|0) $ and $\ket{w}  = (w,1|0)\tran$, and $\Phi$ is the matrix
    \begin{align}
    \Phi & = \left( \begin{array}{cc|c} 0 & 0 & 1 \\ 0 & 0 & 0 \\ \hline 0 & - 1 & 0 \end{array}\right) \, , \label{Phidef}
    \end{align}
    which has the property that if $\bs z = z|\zeta$, $\bs w = w|\psi$ then $D_\zeta \langle \bs z | \bs w \rangle = \bra{\bs z} \Phi \ket{\bs w}$.

  In fact, both the prime form and the Szeg\H o kernel can be seen to arise from the \emph{super prime form} $\mathscr E (\bs z, \bs w)$. If $\srs_g$ is a (not necessarily split) SRS described by a super Schottky group $\bs G$, then we can define \cite{DiVecchia:1988jy}
  \begin{align}
  \mathscr E(\bs z, \bs w) & = (\bs z \dm \bs w) {\prod_{\gamma_\alpha}}' \frac{\bs z \dm \bs\gamma_\alpha(\bs w)}{ \bs z \dm \bs \gamma_\alpha(\bs z)}\frac{\bs w \dm \bs \gamma_\alpha(\bs z)}{\bs w \dm \bs  \gamma_\alpha(\bs w)} \, .
  \end{align}
  We can expand $\log \mathscr E (\bs z , \bs w)$ in $\zeta$ and $\psi$ (where $\bs z = z|\zeta$ and $\bs w = w|\psi$), and if $\srs_g$ is split, then since $\mathscr E$ is even and there are no other odd variables this expansion must have only two terms; in fact
  \begin{align}
  \log \mathscr E(\bs z, \bs w) & = \log E (z,w) \,-\, \zeta\, \psi\,  S(z,w) \, .
  \end{align}

    \subsection{Abelian superdifferentials and the period matrix}
    \label{absfdpm}
    There are super Schottky group expressions for the abelian superdifferentials normalized according to \eq{supabnorm}, and for the period matrix (defined in \eq{srspermatdef}).
    Let us take the Schottky group expression for the bosonic Abelian differentials, \eq{schoab}, expressed in terms of cross-ratios via \eq{abelcross}, then replace the exterior derivative $\d$ with $[\d z | \d \zeta ] D_\zeta$, and replace the Schottky group elements, the fixed points, and the cross-ratio with their SRS equivalents. This gives the super Schotty group expression for the abelian superdifferentials:
    \begin{align}
    \wh \phi_i(z|\zeta) [\d z | \d \zeta] & = [\d z |\d \zeta] \, {\sum_{\bs \gamma_\alpha} }^{(i)} D_\zeta  \log\, \wh \Psi (\bs z , \bs \gamma_\alpha (\bs u_i ), \bs a , \bs \gamma_\alpha (\bs v_i) ) \, , \label{absucr}
    \end{align}
    where the cross-ratio $\wh \Psi$ is defined in \eq{sucr}, the summation is defined in the same way as in \eq{schoab} and $\bs a$ is an arbitrary superpoint on $\mb{CP}^{1|1}$.

    Using the bra-ket notation for $\mb{C}^{2|1}$ introduced in \eq{braket1}, we can write an equivalent expression for \eq{absucr} which is more similar to \eq{schoab} \cite{Magnea:2015fsa}
    \begin{align}
    \wh \phi_i(z|\zeta) [\d z | \d \zeta] & = [\d z |\d \zeta] \, {\sum_{\bs \gamma_\alpha} }^{(i)} \Big(
     \frac{\bra{\bs z} \Phi \bs \gamma_\alpha \ket{\bs u_i}}{\bra{\bs z} \bs \gamma_\alpha \ket{ \bs u_i}} - \frac{\bra{\bs z} \Phi \bs \gamma_\alpha \ket{\bs v_i}}{\bra{\bs z} \bs \gamma_\alpha \ket{ \bs v_i}}
     \Big)\, , \label{absubk}
    \end{align} where $\Phi$ is defined in \eq{Phidef}.

    The expression \eq{absubk} for the abelian superdifferentials allows their periods around the $B_i$ cycles to be computed giving the period matrix $\bs \tau_{ij}$; the result is \cite{DiVecchia:1988jy}
\begin{align}
\bs \tau_{ij} & = \frac{1}{2 \pi \ii} \Big( \delta_{ij} \log \ve_i^2 \,  - \, {\vphantom{\sum_{\bs \gamma_\alpha}}}^{(i)} {\sum_{\bs \gamma_\alpha}}^{'(j)} \log \frac{\bra{\bs u_j} \bs \gamma_\alpha \ket{\bs v_i}}{\bra{\bs u_j}  \bs \gamma_\alpha \ket{ \bs u_i}}\frac{\bra{\bs v_j} \bs \gamma_\alpha \ket{\bs u_i}}{\bra{\bs v_j}  \bs \gamma_\alpha \ket{ \bs v_i}}\Big) \, ,  \label{schopm}
\end{align}
where the summation means the same as in \eq{boschopm}. Using the Cauchy formulas and indefinite integrals given in \S 2.7 of \cite{Friedan:1986rx}, the normalization of the abelian superdifferentials $\wh{\phi}_i({\bs z})$ in \eq{absubk} can be checked, and \eq{schopm} can be derived from \eq{absucr} with a modification of the calculation by Mandelstam in \cite{Mandelstam:1985ww}.

  \subsection{Leading behaviour of expressions in genus $g=2$}
  \label{schoformgtwo}
  The (super) Schottky group formulae given in sections \ref{abdifpm}--\ref{absfdpm} can be used to compute power series in the (semi)multipliers. To a given order in the semimultipliers, only finitely many super Schottky group elements give a non-trivial contribution. The reason is that in all of the above formulae, the contribution from a super Schottky group element $\bs \gamma_\alpha$ is ${\cal O}(\ve_i^n)$ if the reduced word for $\bs \gamma_\alpha$ has at least $n$ powers of the generator $\bs \gamma_i^{\pm 1}$. So, for example, in genus $g=2$, to first order in the semimultipliers we only need to compute the contributions from the 13 elements $\bs \gamma_\alpha \in \{\textbf{Id}, \bs \gamma_1^{\pm1}, \bs \gamma_2^{\pm 1}, \bs\gamma_1^{\pm1}\bs \gamma_2^{\pm1} ,\bs\gamma_2^{\pm1}\bs \gamma_1^{\pm1} \}$.
  In this section, we give the first few terms in this series-expansion of some of the objects in genus $g=2$.
  \subsubsection{Objects defined on the reduced surface}
  In genus $g=2$, there are three Schottky moduli: the one fixed point $u_2 \equiv u$ which is not fixed by \eq{normeq} and the two multipliers $k_i$, $i=1,2$.

  The two abelian differentials can be computed from the expression in \eq{schoab} and are given to leading order in the multipliers $k_i$ by
 \begin{align}
 \omega_1(z)\, \d z & =
 \d z \, \bigg( \frac{1}{z} + \Big( \frac{E_z^2}{u}  + F_z^2 \Big) \, k_2 \, + H (1+u)\bigg( \frac{1}{u} + \frac{1}{z^2} \bigg) \, k_1 \, k_2 \, \bigg)  \, + \, { \cal O}(k_i^2)\nonumber \, 
  \shortintertext{{}}
  \omega_2(z)\, \d z & =
 \d z \, \bigg( \frac{1}{1-z} + \frac{1}{z-u} - (1-u)\bigg( \frac{1}{u} + \frac{1}{z^2} \bigg) \,  k_1  \,  \label{omg2} \shortintertext{{}}
  & \hspace{100pt} - \frac{1-u^2}{u} \Big( \frac{E_z^2}{u} + F_z^2\Big) \,k_1\, k_2 \, \bigg)  + { \cal O}(k_i^2) \, .\nonumber
 \end{align}
 where we have defined the notation
 \begin{align}
 E_z & \equiv \frac{1-u}{1-z} \, ; &
 F_z & \equiv \frac{1-u}{u-z}\, ; &
 H & \equiv \frac{(1-u)^2}{u}   \, . \label{EFH}
 \end{align}
The period matrix $(\tau_{ij})$ can be computed from \eq{schopm}; it is given by:
\begin{align}
(\tau_{ij}) & = \frac{1}{2 \pi \ii } \left( \begin{array}{cc} \log k_1 + 2\,H \, k_2 & \log u -2\,H\,\big( u - \frac{1}{u}\big) k_1 k_2  \\
\log u +2\,H\,\big( u - \frac{1}{u}\big) k_1 k_2 & \log k_2 + 2 \,H \, k_1
\end{array}
\right) + {\cal O}(k_i^2) \, . \label{taug2}
\end{align}
The prime form can be computed from \eq{Ezw}, but for our purpose it is more useful to write down $\partial_z \partial_w \log E(z,w)$ from the formula in \eq{ddlnEzw}. Its leading behaviour in $k_i$ is given by
\begin{align}
\partial_z \partial_w \log E(z,w) & = \frac{1}{(z-w)^2} \, + \, \bigg( \frac{1}{z^2} + \frac{1}{w^2} \bigg) k_1 \, + \, \frac{E_z^2  F_w ^2 + E_w^2 F_z^2}{(1-u)^2} k_2 \,  \label{ddlnEg2}
\shortintertext{{}}
& \hspace{15pt}+ \, \bigg(\frac{E_z^2 + F_z^2}{w^2}+\frac{E_w^2 + F_w^2}{z^2} + F_z^2 + F_w^2 + \frac{E_z^2 + E_w^2}{u^2}\bigg) \, k_1\, k_2  + {\cal O}(k_i^2) \, . \nonumber
\end{align}
To write down the Szeg\H{o} kernel, we should consider the Schottky group as a subgroup of \slc~rather than \psl~so that   $\surf_2$ is a spin curve instead of a plain RS. To choose a spinor bundle $K^{1/2}$, we make a sign choice for the supermultipliers $\ve_i = \ve(\gamma_i)$ of the $g$ generators (this is equivalent to the choice of a $\vartheta$-characteristic $\vec \epsilon_B$ for the $B_i$ cycles; the $\vartheta$-characteristic $\vec \epsilon_A$ for the $A^j$ cycles is always zero with Schottky groups). The formula \eq{szego2} leads to
\begin{align}
S(z,w)
& = \frac{1}{z-w} + \ve_1\Big(1 \,+\, \ve_1 \,- \,H \, \ve_1 \, \ve_2\, (1 + \ve_2)\Big) \bigg( \frac{1}{z} - \frac{1}{w} \bigg) \nonumber \shortintertext{{}}
& \hspace{20pt} + \ve_2 \Big( 1 \,+\, \ve_2 \,-\, H\, \ve_1\, \ve_2 \, (1 + \ve_1)\Big) \frac{E_zF_w - E_w F_z}{1-u}  \label{Szwg2} \shortintertext{{}}
& \hspace{20pt} - \ve_1 \ve_2\big (1 \,+\, \ve_1\, + \,\ve_2\, +\, (1 - H) \,\ve_1 \, \ve_2\big) E_zF_zE_wF_w \frac{(u-wz)^2}{u(1-u)^2} \bigg( \frac{1}{z} - \frac{1}{w} \bigg) \nonumber \, +\,  {\cal O}(\ve_i^3) \, .
\end{align}
  \subsubsection{Canonically normalized supermoduli}
  In this section we write down the first few terms of the abelian superdifferentials and the period matrix for a genus $g=2$ SRS described by a canonically normalized super Schottky group, \ie~one whose two odd supermoduli are both taken to enter via the fixed superpoints of $\bs \gamma_2$, while the odd parameters of $\bs \gamma_1$ are set to zero:
  \begin{align}
  \theta_1 = \phi_1 & = 0 \, ;
  &
  \theta_2 & = \theta \, ;
  &
  \phi_2 & = \phi \, .
  \end{align}
  We can use \eq{absubk} to compute the first few terms of the small-$\ve_i$ expansion of the abelian superdifferentials. Writing $\wh\phi_i(z|\zeta)  = \wh \alpha_i(z) + \zeta\, \wh b_i(z)$, we find
    \begin{align}
    \wh \alpha_1(z) & =  \bigg( \frac{E_z}{u}\, \theta - F_z \phi   + \frac{H}{ z } \bigg(\frac{ \phi}{F_z} -\frac{ \theta}{E_z} \bigg) \, \ve_1 \bigg) \ve_2 + {\cal O}(\ve_i^2) \, , \nonumber
    \shortintertext{{}}
    \wh b_1(z) & = \frac{1}{z} \, + \, \bigg(\frac{E_z^2}{u} + F_z^2 \bigg)\frac{\theta\, \phi}{(1-u)^2} \, \ve_2\,+\, {\cal O}(\ve_i^2) \, ,  \label{absudnorm}
    \shortintertext{and}
    \wh \alpha_2(z) & = \bigg( \frac{ \theta}{u-z} - \frac{ \phi}{1-z} \bigg)\Big(1\, - H \, \ve_1\, \ve_2\Big)   + \bigg(\bigg(\frac{1}{u} - \frac{1}{z}\bigg) \theta - \bigg( 1 - \frac{1}{z}\bigg)\phi \bigg)\, \ve_1 \, +\,  {\cal O}(\ve_i^2)   \, , \nonumber
    \shortintertext{{}}
      \wh b_2(z) & =   \frac{1}{1-z} - \frac{1}{u-z} \, +\,  {\cal O}(\ve_i^2) \, .  \nonumber
    \end{align}
    The period matrix can be computed from \eq{schopm}; it is given by
    \begin{align}
    (\bs \tau_{ij}) & =\frac{1}{2 \pi \ii}  \left( \begin{array}{cc} \log \ve_1^2 \, + \, 2 \frac{1-u}{u} \, \theta \phi \, \ve_2
    &
    \log u \\
    \log u &
    \log \ve_2^2 \, + \, 2 \,  \frac{1-u}{u} \, \theta \phi \, \ve_1 \end{array}\right) \, + \, {\cal O}(\ve_i^2) \, .  \label{pmnormg2}
    \end{align}
  \subsubsection{Supermoduli shared between Schottky generators}
  In this section we write down the first few terms of the abelian superdifferentials and the period matrix for a genus $g=2$ SRS described by a super Schottky group whose two odd supermoduli $\xi$ and $\theta$ enter as in \eq{sharedodd}, parametrizing $\bs \gamma_1$ and   $\bs \gamma_2$ respectively.

   Again using \eq{absubk} to compute the first few terms of the small-$\ve_i$ expansion of the abelian superdifferentials, we write $\wh\phi_i(z|\zeta)  = \wh \alpha_i(z) + \zeta\, \wh b_i(z)$ which gives
    \begin{align}
    \wh \alpha_1(z) & = \, - \,  \frac{\xi}{z}\, + \, \frac{E_z }{u} \Big(\theta\, +\,F_z\, z \, \xi\Big)\ve_2\,+ \, \frac{1-u}{u} \bigg(\bigg(1 - \frac{1}{z}\bigg) \theta \, + \, \frac{1-u}{z} \xi \bigg) \, \ve_1\, \ve_2 \,  + {\cal O}(\ve_i^2) \, , \nonumber
    \shortintertext{{}}
    \wh b_1(z) & =  \frac{1}{z}\, - \,(1-u)\bigg( \frac{\ve_2}{(u-z)^2} \, - \, \frac{\ve_1 \, \ve_2}{u\, z^2} \bigg) \, \theta \xi   \,  + {\cal O}(\ve_i^2) \, \nonumber ,
    \shortintertext{and}
    \wh \alpha_2(z) & = \frac{\theta}{u-z} \, + \, \bigg( \bigg( \frac{1}{u} - \frac{1}{z} \bigg) \theta + \bigg(1 - \frac{1}{u} \bigg) \xi \bigg) \ve_1 \, + \, \frac{H F_z}{1-u} \bigg( E_z \, \xi - \theta \bigg) \, \ve_1 \ve_2\, +\,  {\cal O}(\ve_i^2) \, ,  \nonumber \\
      \wh b_2(z) & =   - \frac{1-u}{(u-z)(1-z)} \, + \, \bigg( \frac{\ve_1}{z^2} - \frac{F_z^2}{u} \, \ve_1 \ve_2 \, \bigg)\, \theta \xi\, +\,  {\cal O}(\ve_i^2) \, .  \label{absudshar}
    \end{align}
    The period matrix with these supermoduli can be computed from \eq{schopm}; it is given by
    \begin{align}
    (\bs \tau_{ij}) & =\frac{1}{2 \pi \ii}  \left( \begin{array}{cc} \log \ve_1^2 \,- \, 2 \frac{1-u}{u} \, \theta \xi \, \ve_2
    &
     \log ( u- \theta \xi) \\
     \log ( u- \theta \xi)  &
    \log \ve_2^2 \, - \, 2 \,  \frac{1-u}{u} \, \theta \xi \, \ve_1 \end{array}\right) \, + \, {\cal O}(\ve_i^2) \, . \label{pmsharedg2}
    \end{align}
    where $\log ( u- \theta \xi) =  \log u \,- \,\frac{1}{u} \, \theta \xi $.
\section{Appendix: Schottky groups and gluing}
\label{SewingApp}
In this appendix we review how (super) Schottky groups for genus $g$ (super) Riemann surfaces can arise from inserting plumbing fixtures between pairs of marked points on lower-genus surfaces, related in a simple way to the (super) Schottky moduli. In Section \ref{bossew} we look at the bosonic case and in Section \ref{supsew} we see how super Schottky groups can arise from gluing pairs of Neveu-Schwarz punctures.

This relation between gluing and Schottky groups is the reason Schottky space integrals arose naturally in early sewing-based computations of multi-loop string amplitudes by Lovelace \cite{Lovelace:1970sj}, Kaku and Yu \cite{Kaku:1970ym} and Alessandrini and Amati \cite{Alessandrini:1971cz,Alessandrini:1971dd}.
\subsection{The bosonic case}
\label{bossew}
\subsubsection{Genus one}
\label{genusonebos}
Let us focus on the construction of a genus $g=1$ Riemann surface in detail. We consider two \emph{a priori} different ways to get a torus from a sphere: firstly by gluing a pair of marked points, and secondly by quotienting a region by a hyperbolic M\"obius map $\gamma$ \emph{\`a la} Schottky, and we will see that the resulting Riemann surfaces are identical if the gluing parameter is identified with the multiplier of $\gamma$.

First of all, consider the Riemann sphere covered by two coordinate charts $z$ and $w$ related by the transition function $w = - 1/z$. Say there are two punctures (\ie~marked points), which can be taken to be at $z=0$ and $w=0$ without loss of generality. In general, given two marked points and two charts $z$, $w$ which vanish at the respective points, we can insert a \emph{plumbing fixture} by identifying
\begin{align}
zw & = - k \label{seweq}
 \end{align}
  on a pair of annular regions (one around each puncture) and removing the discs bounded by the two annuli. The complex parameter $k$ with $|k|< 1$ is called the \emph{gluing parameter}.
 \begin{figure}
\centering
\def\svgwidth{8cm}
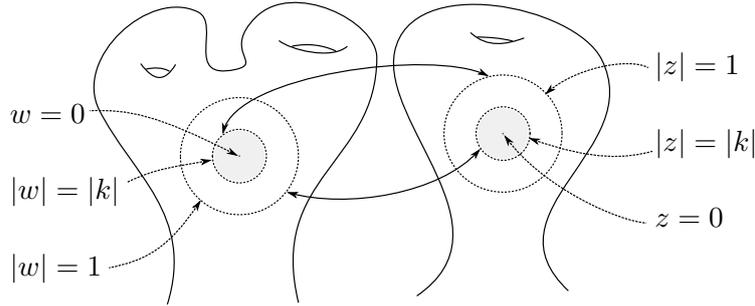
\caption{Gluing a pair of punctures on one or two Riemann surfaces means taking complex coordinate charts $w$, $z$ vanishing at the respective points, cutting the discs $|z| \leq |k|$, $|w|\leq|k|$ from the surface or surfaces and identifying points $P_1$, $P_2$ if $|k|< |w(P_1)| < 1$, $|k|< |z(P_2)|<1$, and $w(P_1) z(P_2) = - k$.
}
\label{fig:sewingsphere}
\end{figure}

In our case, we can turn the sphere into a genus $g=1$ torus by removing the discs $|w| \leq |k|$ and $|z| \leq |k|$ from the sphere and then identify the two annuli $|k| < |z| < 1 $ and $|k| < |w| < 1$ via \eq{seweq}. In fact, since we have removed the point $w=0$, we can describe everything in terms of the $z$ coordinate, so the second annulus is given by $1< |z| < |k|^{-1}$. But in gluing, we have stipulated that every point in the first annulus is equivalent to a point in the second annulus {via} \eq{seweq}, so every point on the torus we have built (except points with $|z|=1$) lies under \emph{two} points on the annulus $|k| < |z| < |k|^{-1}$, with the two $z$ coordinates of those points differing from each other by a factor of $k$. So half of the annulus is superfluous, and the resulting torus can be described as an annulus whose inner and outer radii differ by by a factor of $|k|$, where two points $P_{\,\text{i}}$ and $P_{\text{o}}$ on the inner and outer boundaries are identified if $z(P_{\,\text{i}}) = k z(P_{\text{o}})$ (see \Fig{fig:sewingsphere}).  $k$ is the modulus in the resulting family of tori.

\begin{figure}
\begin{center}
\centering
\def\svgwidth{5cm}
\subfloat[]{ 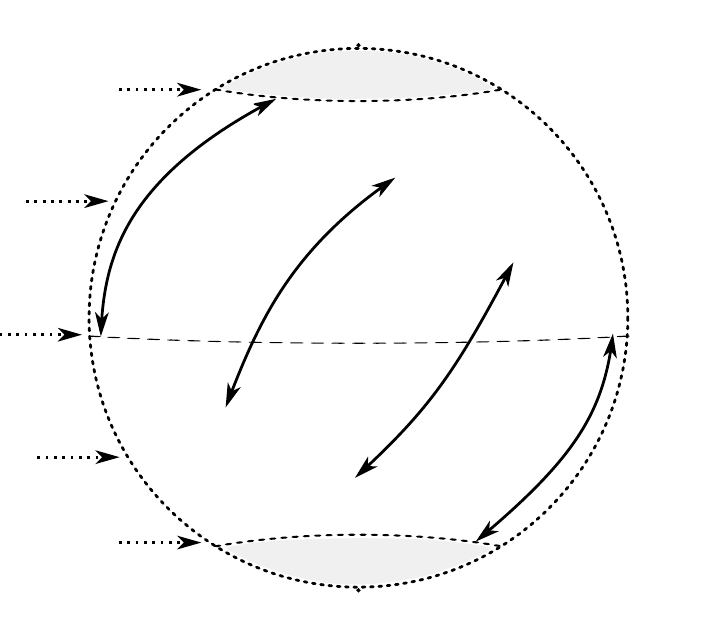 \label{fig:sewingsphere} }
\def\svgwidth{5cm}
\subfloat[]{ 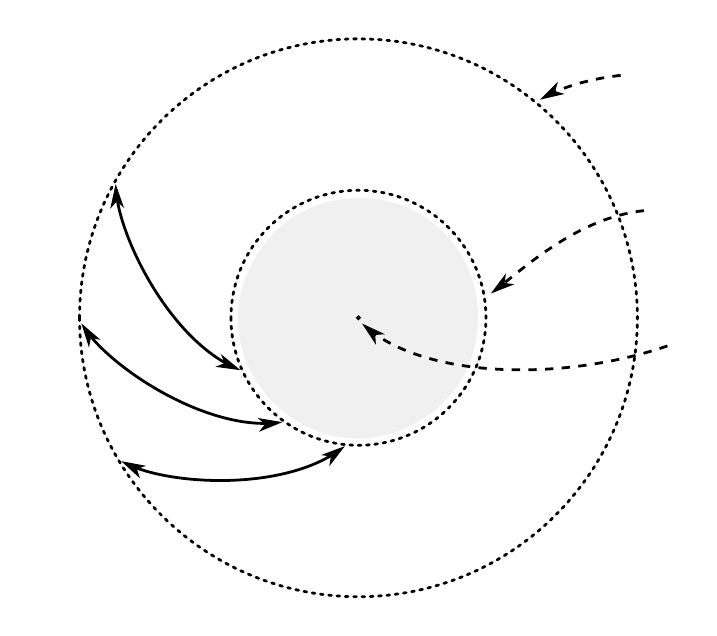 \label{fig:annulusdomain} }
\caption{Inserting a plumbing fixture between two points on a sphere is equivalent to identifying annuli around the two points and removing the discs at their middles. We end up with an annulus (\Fig{fig:annulusdomain}) whose inner and outer boundaries are identified. Solid arrows are drawn to indicate some of the pairs of points which are to be identified. \label{fig:genusone}}
\end{center}
\end{figure}

Now let us describe the Schottky group construction of a torus. Consider a single {hyperbolic} M\"obius map $\gamma \in \psl$, so $\gamma$ has two distinct fixed points, one attractive and one repulsive. We can choose a coordinate $z$ for the sphere in which these fixed points are at $z=0$ and $z = \infty$, respectively, so $\gamma$ is a dilatation with multiplier $k$, $\gamma = P_k$ as defined in \eq{Pkdef}.

To build a torus \emph{\`a la} Schottky, we remove the two fixed points (which constitute the whole limit set $\Lambda(G)$ in genus one) from the sphere and quotient what is left by the group generated by $\gamma$, \ie~the group of integer powers of $P_k$. In terms of the $z$ coordinates, this means removing $z=0$ and $z=\infty$ and quotienting by the equivalence relation
\begin{align}
P_1 \sim P_2 & \leftrightarrow z(P_1) = k^n z(P_2) \text{ for some } n \in \mb{Z} \, .
\label{torusschottky}
\end{align}
If we take an annular region bounded by two circles centred at $z=0$ whose radii differ by a factor of $|k|$, then every point in $\Omega_1 \equiv \rs - \{z=0,z = \infty\}$ is equivalent to either a point in the interior of this annulus or a point in each of the two boundary components. The annulus, therefore, constitutes a fundamental region for the action of $P_k$ on $\Omega_1$, and the torus resulting from the quotient can be constructed as the annulus whose two boundary components are identified as before.

\begin{figure}
\begin{center}
\centering
\def\svgwidth{5cm}
 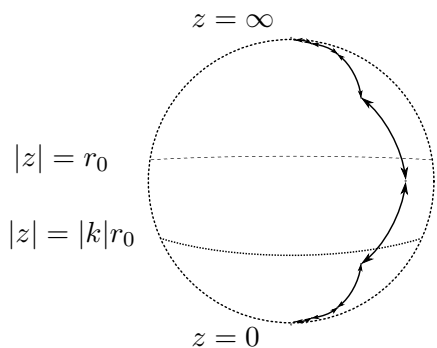 \label{fig:schottkytorus}
\caption{The Schottky group description of a genus one surface may be achieved by subtracting the points $z=0$ and $z=\infty$ from $\rs$ and imposing the equivalence relation $z \sim k^n z$, $n \in \mb{Z}$. $k$ is the modulus. The solid arrows indicate one equivalence class of points. The band $|k|r_0 \leq |z| < r_0$ is a fundamental region for the group action. The resulting quotient space is equivalent as a Riemann surface to the one obtained by gluing in \Fig{fig:genusone}.}
\end{center}
\end{figure}
We see, therefore, that gluing a pair of punctures on the sphere gives the same Riemann surface as quotienting a region $\Omega_1$ by a hyperbolic M\"obius map, so building a torus \emph{\`a la} Schottky is one way to effect a gluing, with the Schottky modulus $k$ identified with the gluing parameter.

Of course, nothing is special about the points $z=0$ and $z=\infty$. Punctures at two arbitrary points $z=u$ and $z=v$ can be sewn together with gluing parameter $k$ if we quotient $(\rs - \{ u,v\})$ by the hyperbolic M\"obius map $\gamma$ defined in \eq{gamuvk}, since a change of coordinates by $\Gamma_{uv}$ as defined in \eq{Gamuvdef} transforms this to the case which has already been dealt with.

\subsubsection{Higher genus Riemann surfaces}
\label{sewhigherg}
The construction of compact Riemann surfaces of genus $g$ is described in Section \ref{schothigherg}. In fact, any Schottky group $G$ can be seen to arise by gluing pairs of punctures located at the fixed points in some sense, as we saw in the genus $g=1$ case in the previous section.

This can be seen inductively. Consider a genus $g$ surface $\surf_g$ defined by a Schottky group $G = \langle \gamma_1, \ldots, \gamma_g \rangle$. Then the rank-$(g-1)$ group $\wt G = \langle \gamma_1 , \ldots , \gamma_{g-1}\rangle$ is a Schottky group defining a genus $(g-1)$ surface $\surf_{g-1}$. Any fundamental region ${\cal F}$ for $G$ can have added to it the discs bounded by the Schottky circles ${\cal C}_g$, ${\cal C}_g'$ to yield a fundamental region $\wt{\cal F}$ for $\wt G$, with the fixed points $u_g$ and $v_g$ of $\gamma_g$ in the interior of $\wt{\cal F}$. The equivalence classes $\wt{G}(u_g)$ and $\wt{G}(v_g)$ correspond to a pair of marked points on $\surf_{g-1}$.

Now, if we did not have to worry about $\wt{G}$ then as discussed at the end of Section \ref{genusonebos}, quotienting by $\gamma_g$ would amount to gluing a pair of punctures at $z=u_g$ and $z=v_g$ with gluing parameter $k_g$. But now this has to be done in a way preserving invariance under the rank-$(g-1)$ Schottky group $\wt G$. Hence we must quotient an appropriate covering space by all M\"obius maps in the original rank-$g$ Schottky group $G = \langle \wt{G} , \gamma_g \rangle$. Thus, the Schottky modulus $k_g$ can be thought of as a gluing parameter for the $g$'th handle. Of course, the same argument could apply to any primitive Schottky group element $\gamma_\alpha$, and in particular the $g$ generators $\gamma_1 , \ldots , \gamma_g$ of a marked rank-$g$ Schottky group could be associated with $g$ gluings needed to build a genus-$g$ Riemann surface $\surf_g$.

\begin{figure}
\begin{center}
\centering
\def\svgwidth{3cm}
\subfloat[]{ 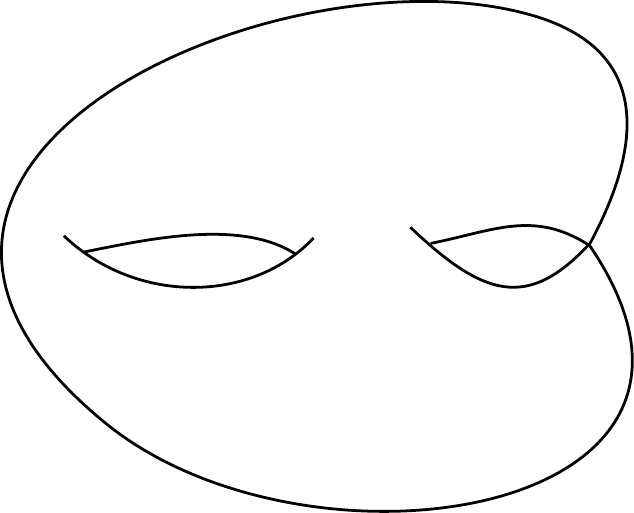 \label{fig:nonsep} }
\def\svgwidth{6cm}
\subfloat[]{ 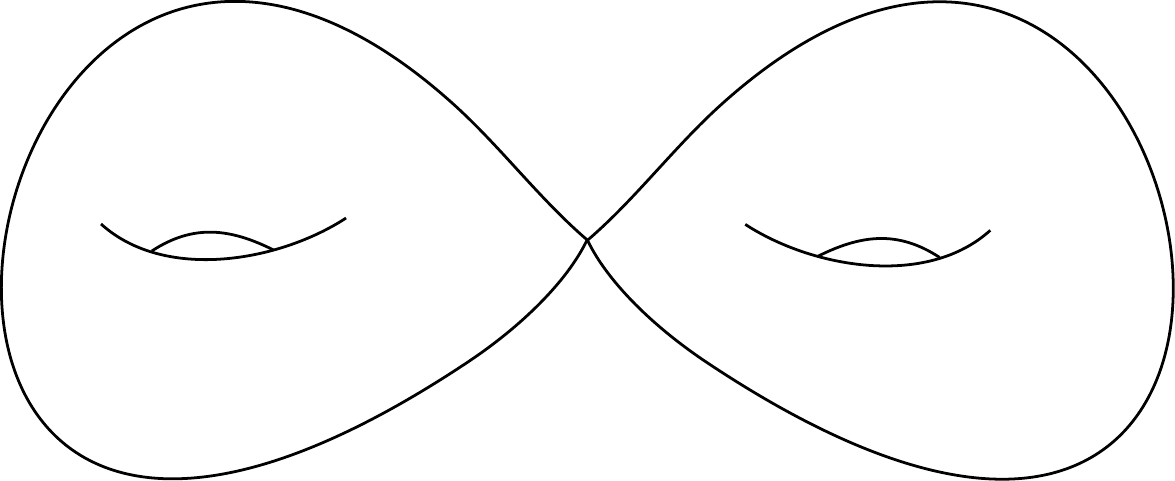 \label{fig:sep} }
\caption{Non-separating (\Fig{fig:nonsep}) and separating (\Fig{fig:sep}) degenerations of a genus $g=2$ surface. \label{fig:degens}}
\end{center}
\end{figure}

The Deligne-Mumford compactification of moduli space $\overline{\cal M}_g$ is obtained by extending families of surfaces which can locally be described with plumbing fixtures $xy=-k$ to include the limiting surface as $k\to 0$; these surfaces are degenerate with double points. Degenerations can be classified topologically according to whether they are \emph{separating} or \emph{non-separating}, \ie~whether the degenerate surface is the disjoint union of two surfaces with one extra marked point on each, or a single surface with two marked points (see \Fig{fig:degens}).
  Here we describe two ways in which points on the compactification divisor $\EuScript{D}_g = \overline{\cal M}_g - {\cal M}_g$ can be reached in Schottky group language. Firstly, according to the discussion in the previous section, the multiplier $k_\alpha$ of any primitive $\gamma_\alpha \in G$  is a gluing parameter associated to some handle, so if we let it vanish, $k_\alpha \to 0$, we get a ``non-separating'' degeneration in which a non-trivial homology cycle is pinched. For the multipliers $k_i$ of the $g$ Schottky generators $\gamma_i$, the limit $k_i \to 0$ describes the degeneration in which the $A^i$ homology cycle pinches off.

  To describe some separating degenerations we need to do a bit more work. Given two compact surfaces $\surf_{g_L}$ and $\surf_{g_R}$ of genus $g_L$ and $g_R$, respectively, described by Schottky groups, we can glue the surfaces together to obtain a new surface $\surf_{g_L+g_R} = \surf_{g_L} \# \surf_{g_R}$ in the following way.

\begin{figure}
\begin{center}
\centering
\subfloat[]{ 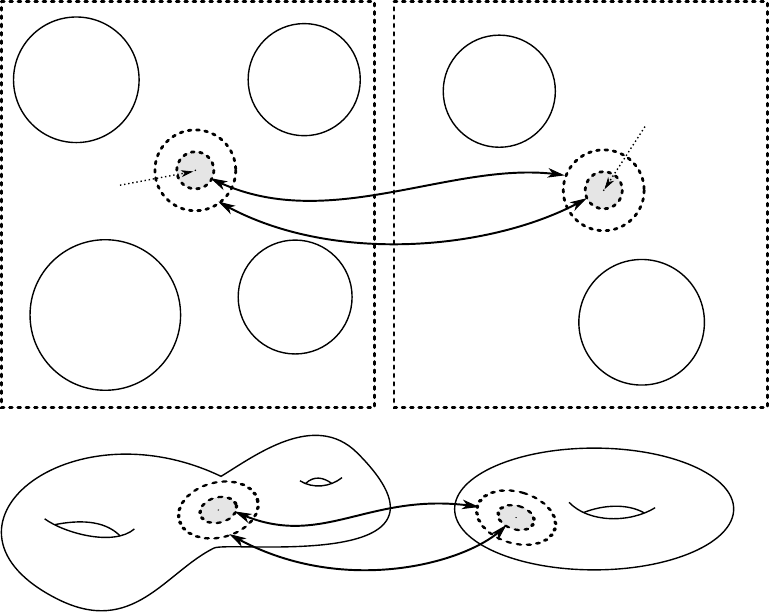 \label{fig:sss} }
\subfloat[]{ 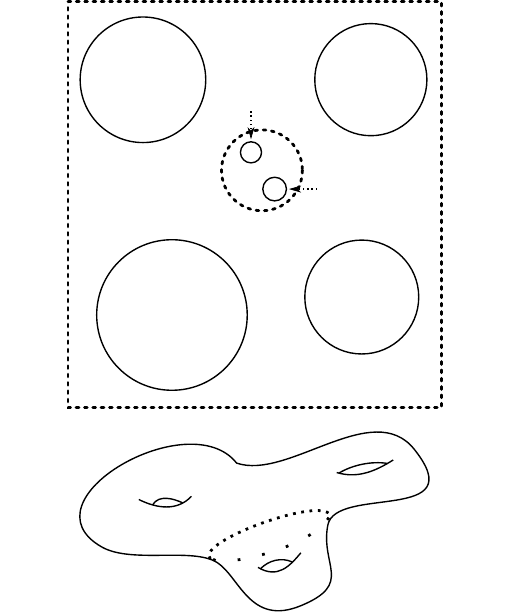 \label{fig:sse} }
\caption{Two surfaces $\surf_L$, $\surf_R$ described by Schottky groups $G_L$, $G_R$ (shown in \Fig{fig:sss} for groups with rank 2 and 1, respectively) can be `glued' together at a pair of punctures using Eqs.~\ref{wtzwdef} -- \ref{zgendef}. This gives a higher genus surface described by the Schottky group $G = \langle G_L, \widetilde{\Gamma}G_R\widetilde{\Gamma}^{-1}   \rangle $ (an example is shown in \Fig{fig:sse}: the new Schottky circles ${\cal C}_3$ and ${\cal C}_3'$ are given by $\widetilde\Gamma \widetilde{\cal C}_1$ and $\widetilde\Gamma \widetilde{\cal C}_1'$ respectively).}
\label{fig:connsum}
\end{center}
\end{figure}

Let the first surface $\surf_{g_L}$ be described by a uniformizing coordinate $z$ quotiented by the Schottky group $\sg_L$ generated by $\gamma_i$, $i=1,\ldots , g_1$, and the second surface described by a uniformizing coordinate $w$ quotiented by a Schottky group $\sg_R$ generated by $\wt \gamma_j$, $j=1, \ldots, g_R$.  Say we want to put a plumbing fixture between the point $z=z_0$ on $\surf_{g_L}$ and the point $w=w_0$ on $\surf_{g_R}$. Then we need to find local coordinates vanishing at these two marked points; we can take
\begin{align}
\wt{z} & = \frac{z-z_0}{z-z_1} \equiv f_L(z) \, ,
&
\wt{w} & = \frac{w-w_0}{w-w_1} \equiv f_R(w) \, , \label{wtzwdef}
\end{align}
where $z_1$ and $w_1$ are free parameters. Then to glue with gluing parameter $q$,  we identify
\begin{align}
\wt{z} \wt{w} & = - q \, \label{ztwtq}
\end{align}
on annuli around the two punctures. Using \eq{wtzwdef}, this gives a relationship between the $z$ and $w$ coordinates,
\begin{align}
z & = \frac{(qz_1+z_0) w - (qz_1 w_1 + z_0 w_0)}{(1+q)w - (w_0 + q w_1)} \equiv \wt{\Gamma}(w) \, . \label{Gamwtoz}
\end{align}
Since this is a M\"obius map, we can use it to define a new coordinate system everywhere on the Schottky cover of $\surf_{g_R}$; in particular, we can find $z$ coordinates for all of the fixed points, and replace the Schottky group generators on $\Sigma_{g_R}$ with
\begin{align}
z & \mapsto \gamma_{g_L+j}(z) \equiv ( \,  \wt{\Gamma} \circ \wt{\gamma}_j \circ \wt{\Gamma}^{-1} \, ) (z) \, & j & = 1 , \ldots , g_R \, . \label{zgendef}
\end{align}
The group
\begin{align}
\sg & = \langle \gamma_i, \gamma_{g_L + j } ; \,  i=1 , \ldots , g_L ; \, j = 1 , \ldots g_R \rangle   \,  \label{newshotgens}
\end{align}
is a rank-$(g_L +g_R)$ Schottky group for sufficiently small values of the gluing parameter $q$,
and the surface this group describes is exactly the surface obtained by gluing the two lower genus surfaces in the above way. The $3(g_L + g_R) -3$ moduli come from the $(3g_L - 3) + (3g_R - 3)$ moduli of the two sewn surfaces $\surf_{g_L}$, $\surf_{g_R}$ plus the 3 complex parameters of the gluing map $\widetilde \Gamma$ in \eq{Gamwtoz} (see \fig{fig:connsum}).

If we can present a rank-$(g_L+g_R)$ Schottky group as in \eq{zgendef} and \eq{newshotgens}, then allowing the gluing parameter to vanish $q \to 0$ corresponds to the separating degeneration in which a genus-$(g_L+g_R)$ surface $\surf_{g_L+g_R}$ pinches off into two surfaces of genus $g_L$ and $g_R$ connected at a node.

Not all of $\overline{\mathcal{M}}_g$ is easily described in Schottky language; for example, there is not a convenient way to describe degenerations which pinch the $B_i$ homology cycles.

\subsection{Super Schottky groups and gluing}
\label{supsew}
\subsubsection{Building a genus one SRS by gluing}
 As for Riemann surfaces, there is a notion of gluing pairs of punctures on super-Riemann surfaces. In fact, there are two different types of punctures: \emph{Neveu-Schwarz} (NS) and \emph{Ramond} (R) \cite{Cohn:1987cj}. An NS puncture may be thought of as a marked point on the SRS, given by $z|\zeta = z_0 | \zeta_0$ in some coordinate chart. An R puncture is something quite different, characterized by a singularity in the superconformal structure of \surf. At an $R$ puncture, a section $D$ of ${\cal D}$ fails to be linearly independent of $D^2$. In other words, near an R puncture there are some (non-superconformal) coordinates $y|\psi$ in which ${\cal D}$ has sections spanned by $D_\psi^* = \partial_\psi + \psi y \partial_y$, whose square $(D_\psi^*)^2 = y \partial_y$ vanishes at $y=0$ \cite{Witten:2012ga}. There are notions of gluing pairs of punctures of either NS or R type, but in this paper we will only consider gluing pairs of NS punctures.

 Our immediate goal is to see how, just as in the bosonic case, we can build a higher-genus SRS by gluing pairs of NS punctures on $\mb{CP}^{1|1}$ in a way that is equivalent to quotienting a subdomain of $\mb{CP}^{1|1}$ by a super Schottky group.

 Given two superconformal coordinate charts $z|\zeta$ and $w|\psi$ which vanish respectively at a pair of NS punctures on a SRS, an NS-gluing of the punctures gives a family of SRS by imposing a relation (Eq.~(6.9) of \cite{Witten:2012ga})
 \begin{align}
 z w & = - \varepsilon^2 \, ,  & z \psi  & = \varepsilon \zeta \, ,  &  w \zeta & = - \varepsilon \psi \, , & \psi \zeta & = 0 \, \label{NSsew}
 \end{align}
 on a pair of regions whose reduced spaces are annuli (one around each puncture), and removing the discs inside each annulus.
 Here $\varepsilon$ is an even parameter with $|\ve| < 1$ called the \emph{NS gluing parameter}.\footnote{When we refer to some expression's absolute value $|\cdot|$ in an SRS context, we always mean modulo any odd variables.} Note that the first equation of \eq{NSsew} matches \eq{seweq} describing gluing of a pair of punctures on an Riemann surface if $\varepsilon^2  = k$.

 Let us begin by considering a pair of NS punctures on $\mb{CP}^{1|1}$, then we can use \osp~automorphisms to assume without loss of generality that the punctures are at $z|\zeta = 0|0$ and $w|\psi = 0|0$ in terms of the coordinates introduced in \eq{affinecharts}. Then just as in the bosonic case, we can remove the discs $|w| \leq |\varepsilon^2|$ and $|z| \leq |\varepsilon^2|$ from the surface and identify the two annuli $|\varepsilon^2| < |z| < 1 $ and $|\varepsilon^2| < |w| < 1$ {via} \eq{NSsew}.

 Again, since we have removed the locus $w=0$, we should be able to describe everything in terms of the $z|\zeta$ superconformal coordinates, where the second annulus mentioned above is given equivalently by $1< |z| < |\varepsilon|^{-2}$. But in gluing, we have stipulated that every super-point in the first annulus is equivalent to a super-point in the second annulus {via} \eq{NSsew}, so (apart from those points with $|z|=1$) every super-point on the genus one SRS which we have built lies under two distinct super-points $\bs{z}_1=z_1|\zeta_1$ and $\bs{z}_2=z_2|\zeta_2$ on the annular region defined by $|\varepsilon|^2 < |z| < |\varepsilon|^{-2}$. For a given super-point $\bs{z}_1$ with $1 < |z_1| < |\varepsilon|^{-2}$, we can first use \eq{affinecharts} to find its $w|\psi$ coordinates, and then use \eq{NSsew} to write down the $z|\zeta$ coordinates of its corresponding point in the annulus $|\varepsilon|^2 < |z| < 1$, giving us $\bs{z}_2$. We find that the first and second equations of \eq{NSsew} lead, respectively, to
 \begin{align}
 z_2 & = \varepsilon^2\, z_1 \, , & \zeta_2 & = \varepsilon \, \zeta_1 \, , \label{Z1Z2}
 \end{align}
 with the third equation of \eq{NSsew} being interchangeable, here, with the second one, and the fourth being satisfied since $\psi \propto \zeta_1 \propto \zeta_2$.

   So just as before (see \Fig{fig:sewingsphere}), half of the annulus is superfluous, and the resulting genus one SRS can be described as an annular region in $\mb{C}^{1|1}$ whose inner and outer radii differ by by a factor of $|\varepsilon^2|$, where two super-points on the outer and inner boundaries with coordinates $\bs{z}_1$ and $\bs{z}_2$ are identified if \eq{Z1Z2} holds.  $\varepsilon$ is the modulus in the resulting family of SRS.

From this point onwards, we will fix the superconformal chart $\bs{z} = z|\psi$ defined in \eq{affinecharts} as a canonical chart on $\mb{CP}^{1|1}$, writing the point $\bs w = 0|0$ as $\bs z = \bs \infty$. Also, denote $\bs 0 \equiv 0|0$.

   There is also a Schottky group construction of the same SRS, and we will see as in section \ref{genusonebos} that some of the Schottky moduli are naturally identified with NS gluing parameters.
   Notice that the equivalence in \eq{Z1Z2} may be rewritten in terms of $\bs{P}_\ve$ as
    \begin{align}
    \bs{z}_2 & = \bs{P}_\ve(\bs{z}_1) \, .
    \end{align}
The Schottky construction of a genus one SRS proceeds similarly to the bosonic case. The two fixed superpoints of a single generator $\bs{\gamma}$ are removed from $\mb{CP}^{1|1}$ to get the Schottky cover SRS, which is then quotiented by the action of $\bs{\gamma}$. With only one generator we can take $\bs{\gamma} = \bs{P}_\ve$ without loss of generality, then we find that the resulting quotient space can be described as an annular fundamental region in $\mb{C}^{1|1}$ whose inner and outer radii differ by by a factor of $|\varepsilon^2|$, where two super-points $\bs{z}_1$ and $\bs{z}_2$ on the outer and inner boundaries are identified if \eq{Z1Z2} holds. This shows that building a genus one SRS with a super Schottky group characterized by a semimultiplier $\varepsilon$ is exactly the same as NS-gluing a pair of NS punctures on $\mb{CP}^{1|1}$, with $\varepsilon$ as the NS gluing parameter.

   To see the relationship between this construction of a genus one SRS and the  more common one in terms of quotienting $\mb{CP}^{1|1}$ by a $\mb{Z}\times\mb{Z}$ lattice (\cf~section 5.2 of \cite{Witten:2012ga}), consider a new superconformal coordinate $\bs{r} \equiv r|\rho$ defined by
   \begin{align}
   z & = \ex{r} \, , & \zeta & = \ex{r/2} \rho \, . \label{zTor}
   \end{align}
   Then the transformation $\bs{P}_\ve$ of \eq{zPe} acts on these new coordinates via
   \begin{align}
   r & \mapsto r + \log \ve^2 \, , & \rho & \mapsto \rho \, . \label{lattice1}
   \end{align}
   Also, since \eq{zTor} means that $r+2 \pi \ii$ corresponds to the same value of $z$ as $r$ does, we need to quotient by the group generated by
   \begin{align}
   r & \mapsto r + 2 \pi \ii \, , & \rho & \mapsto - \rho \, , \label{lattice2}
   \end{align}
   where the minus sign in the second equation comes from single-valuedness of $\zeta$ in \eq{zTor}, and is crucial since it means that the super-torus obtained by quotienting $\mb{C}^{1|1}$ by \eq{lattice1} and \eq{lattice2} has an \emph{even} spin structure.  \eq{lattice2} means that in the genus one case, we have $\epsilon_a = 0$, and by similar arguments $\vec\epsilon_a = \vec 0$ for higher genus super-Schottky group constructions, which is why we can only describe SRS of even spin structure in this way \cite{Manin1991}.

  \subsubsection{Higher genus super-Riemann surfaces}
   The construction of genus-$g$ SRS with super Schottky groups can be found inductively by gluing pairs of NS punctures on genus-$(g-1)$ SRS, requiring compatibility with the rank-$(g-1)$ super Schottky group. This proceeds analogously to the construction of higher genus Riemann surfaces \emph{via}~Schottky groups in section \ref{sewhigherg}, and leads to the construction described in section \ref{ssg}.

   Each of the $g$ semimultipliers $\varepsilon_i$ is an NS gluing parameter. The Deligne-Mumford compactification of super-Moduli space includes the loci  where $\varepsilon_i \to 0$, which correspond topologically to pinching off of the $A^i$ homology cycle on the underlying surface.

   Just as in the bosonic case, we can describe the gluing together of two compact SRS concretely in terms of super Schottky moduli. Let $\srs_{g_L}$ and $\srs_{g_R}$ be two compact SRS of genus $g_I$ described by Schottky groups $\bs{G}_I$ for $I=L,R$. Suppose we have one NS puncture $P_I$ on each surface, lying under the equivalence class $\bs{G}_I(\bs{p}_I) \subseteq \bs{\Omega}(\bs{G}_I)$, where $\bs{p}_I$ is any representative of $P_I$ in the super Schottky cover. Let us say the superconformal coordinates on $\bs{\Omega}(\bs{G}_L)$ and $\bs{\Omega}(\bs{G}_R)$ are  $\bs{z}$ and $\bs{w}$, respectively. On each super Schottky cover $\bs{\Omega}(\bs{G}_I)$ let us make \osp~changes of coordinates $\wt{\bs{z}} = \bs{f}_L(\bs{z})$, $\wt{\bs{w}} = \bs{f}_R(\bs{w})$ such that $\bs{f}_I(\bs{p}_I) = \bs{0}$. Then the two NS punctures may be glued together with NS gluing parameter $r$ according to \eq{NSsew} if we impose the following identification:
   \begin{align}
   \wt{\bs{w}} & = (\bs{I} \circ \bs{P}_r^{-1}) (\wt{\bs{z}}) \label{sepsew}
   \end{align}
   where $\bs{I}$ and $\bs{P}_r$ are defined in \eq{ztoy} and \eq{spdef}, respectively. Using this identification between the coordinates on the two super Schottky covers, we can find a super-Schottky group $\wt{\bs{G}}_R$ conjugate to $\bs{G}_R$ which generates, together with $\bs{G}_L$, the rank-$(g_L+g_R)$ group
   \begin{align}
    \bs G & = \big\langle \bs G_L , \, \, \wt{\bs{\Gamma}} ^{-1}  \bs G_R  \wt{\bs{\Gamma}} \big\rangle \, , &  \wt{\bs{\Gamma}}   & =  \bs{f}_R^{-1} \bs{I}  \bs{P}_r^{-1} \bs{f}_L\, , \label{Gsrssep}
   \end{align}
   which is a super Schottky group for sufficiently small values of $r$.
    The Deligne-Mumford compactification of super moduli space includes the limiting SRS as $r\to 0$, which corresponds topologically to $\srs_{g_L + g_R}$ pinching off into two surfaces $\srs_{g_L}$ and $\srs_{g_R}$ joined at an NS node.

    As an example, let us compute the behaviour of the off-diagonal elements of the period matrix ${\bs \tau}_{ij}$ of a non-split genus $g=2$ SRS $\srs_2$ near a separating degeneration in super-moduli space. Let us build $\srs_2$ by gluing together two genus $g=1$ SRS, $\srs_L$ and ${\srs}_R$, in the above manner. Let $\srs_L$ and $\srs_R$ be described by the rank-1 super Schottky groups ${\bs G}_L$ and ${\bs G}_R$ generated by ${\bs \gamma}_{1} = {\bs P}_{\ve_1}$ and ${\bs \gamma}_2 = {\bs P}_{\ve_2}$, respectively, where the $\ve_i$ are the even supermoduli of the two surfaces. Let the super Schottky cover of $\srs_L$ have superconformal coordinates $\bs z$, with a NS puncture at $z|\zeta = 1 | \alpha$ (and all its images under ${\bs G}_L$). Similarly let  the super Schottky cover of $\srs_R$ have superconformal coordinates $\bs w$, with a NS puncture at the ${\bs G}_R$-equivalence class of $w|\psi = 1 | \beta$. Then the maps ${\bs f}_I$ are given by
    \begin{align}
  {\bs f}_L & = \left(\begin{array}{cc|c} 1 & -1 & \alpha \\ 0 & 1 & 0 \\ \hline 0 & - \alpha & 1 \end{array}\right) \, ,
  &
  {\bs f}_R & = \left(\begin{array}{cc|c} 1 & -1 & \beta \\ 0 & 1 & 0 \\ \hline 0 & - \beta & 1 \end{array}\right) ,
    \end{align}
    which (along with ${\bs P}_r$ from \eq{zPe} and ${\bs I}$ from \eq{ztoy}) leads to the following expression for $\wt{\bs \Gamma}$ as defined in \eq{Gsrssep}:
    \begin{align}
    \wt{\bs \Gamma} & = \left(\begin{array}{cc|c} - {1}/{r} & {1}/{r} + r - \alpha \beta & - {\alpha}/{r} - \beta \\
    - {1}/{r} &{1}/{r} & - {\alpha}/{r} \\ \hline
    - {\beta}/{r} & - \alpha + {\beta}/{r} & 1 + {\alpha\beta}/{r} \end{array}\right) \, .
    \end{align}
    Now, this is exactly of the form of ${\bs \Gamma}_{\bs u \bs v}$ in \eq{Gammauvdef} if we set
    \begin{align}
    u|\theta & = 1 + r^2 - r \alpha \beta \, \big| \, \alpha + r \beta &
    v|\phi  &= 1 \, \big| \, \alpha \, ,
    \end{align}
    and thus the super Schottky group for the sewn surface is ${\bs G} = \langle {\bs P}_{\ve_1} , {\bs \Gamma}_{\bs u \bs v} ^{-1} {\bs P}_{\ve_2} {\bs \Gamma}_{\bs u \bs v} \rangle$. This matches the canonically normalized super Schottky group in genus $g=2$ \eq{canonsc} if we take the NS-gluing supermoduli to be related to the canonical Schottky supermoduli by:
    \begin{align}
 \alpha & = \phi \, , & \beta & = \frac{\theta - \phi}{r} \, , &    r^2 & = -(1 - u + \theta \phi) \, . \label{degenscho}
    \end{align}
    Now, we have already seen (\eq{pmnormg2}) that the off-diagonal term of the period matrix is given, to lowest order in the super Schottky semimultipliers $\ve_i$, by ${\bs \tau}_{12} = \frac{1}{2 \pi \ii} \log u + {\cal O}(\ve_i^2)$. In superstring perturbation theory, it is the NS gluing parameter $r$ which must be held fixed while integrating over the odd supermoduli in a neighbourhood of a degeneration. A computation of the 2-loop superstring amplitude based on integrating over the odd supermoduli with fixed period matrix entries ${\bs \tau}_{ij}$ is described by D'Hoker and Phong in \cite{D'Hoker:2002gw}. The fact that the NS gluing parameter $r$ in \eq{degenscho} is a function not just of $u$ but also $\theta$ and $\phi$ is another way to see that such a procedure needs a correction corresponding to the separating degeneration near $\bs \tau_{12} \to 0$, as explained by Witten in section 3 of \cite{Witten:2013cia}.
\bibliographystyle{jhep}
\bibliography{rs}

\providecommand{\href}[2]{#2}\begingroup\raggedright\begin{thebibliography}{10}

\bibitem{Witten:2012ga}
E.~Witten, {\it {Notes On Super Riemann Surfaces And Their Moduli}},
  \href{http://xxx.lanl.gov/abs/1209.2459}{{\tt 1209.2459}}.

\bibitem{Donagi:2013dua}
R.~Donagi and E.~Witten, {\it {Supermoduli Space Is Not Projected}},
  \href{http://xxx.lanl.gov/abs/1304.7798}{{\tt 1304.7798}}.

\bibitem{Donagi:2014hza}
R.~Donagi and E.~Witten, {\it {Super Atiyah classes and obstructions to
  splitting of supermoduli space}},
  \href{http://xxx.lanl.gov/abs/1404.6257}{{\tt 1404.6257}}.

\bibitem{Jost:2014wfa}
J.~Jost, E.~Keßler, and J.~Tolksdorf, {\it {Super Riemann surfaces, metrics,
  and the gravitino}},  \href{http://xxx.lanl.gov/abs/1412.5146}{{\tt
  1412.5146}}.

\bibitem{Witten:2015hwa}
E.~Witten, {\it {The Super Period Matrix With Ramond Punctures}},  {\em J.
  Geom. Phys.} {\bf 92} (2015) 210--239,
  [\href{http://xxx.lanl.gov/abs/1501.02499}{{\tt 1501.02499}}].

\bibitem{D'Hoker:2015kwa}
E.~D'Hoker and D.~H. Phong, {\it {The Super Period Matrix with Ramond Punctures
  in the supergravity formulation}},  {\em Nucl. Phys.} {\bf B899} (2015)
  772--809, [\href{http://xxx.lanl.gov/abs/1501.02675}{{\tt 1501.02675}}].

\bibitem{Friedan:1986rx}
D.~Friedan. \emph{Notes on String Theory and Two Dimensional Conformal Field
  Theory}, pp.~162--213 in M.~B.~Green \emph{et al.}, eds., \emph{Unified
  String Theories} (World Scientific, 1986).

\bibitem{Friedan:1985ge}
D.~Friedan, E.~J. Martinec, and S.~H. Shenker, {\it {Conformal Invariance,
  Supersymmetry and String Theory}},  {\em Nucl. Phys.} {\bf B271} (1986) 93.

\bibitem{D'Hoker:1988ta}
E.~D'Hoker and D.~Phong, {\it {The Geometry of String Perturbation Theory}},
  {\em Rev.Mod.Phys.} {\bf 60} (1988) 917.

\bibitem{Witten:2012bh}
E.~Witten, {\it {Superstring Perturbation Theory Revisited}},
  \href{http://xxx.lanl.gov/abs/1209.5461}{{\tt 1209.5461}}.

\bibitem{Witten:2013cia}
E.~Witten, {\it {More On Superstring Perturbation Theory}},
  \href{http://xxx.lanl.gov/abs/1304.2832}{{\tt 1304.2832}}.

\bibitem{Witten:2013tpa}
E.~Witten, {\it {Notes On Holomorphic String And Superstring Theory Measures Of
  Low Genus}},  \href{http://xxx.lanl.gov/abs/1306.3621}{{\tt 1306.3621}}.

\bibitem{D'Hoker:1989ai}
E.~D'Hoker and D.~Phong, {\it {Conformal Scalar Fields and Chiral Splitting on
  Superriemann Surfaces}},  {\em Commun.Math.Phys.} {\bf 125} (1989) 469.

\bibitem{D'Hoker:2014nna}
E.~D'Hoker, {\it {Topics in Two-Loop Superstring Perturbation Theory}},
  \href{http://xxx.lanl.gov/abs/1403.5494}{{\tt 1403.5494}}.

\bibitem{Martinec:1986bq}
E.~J. Martinec, {\it {Conformal Field Theory on a (Super)Riemann Surface}},
  {\em Nucl.Phys.} {\bf B281} (1987) 157.

\bibitem{Manin1986}
Y.~I. Manin. \emph{Quantum Strings and Algebraic Curves}, pp.~1286--1295 in
  A.~M.~Gleason, ed., \emph{Proceedings of the International Congress of
  Mathematicians, Berkeley, California, August 3--11, 1986} (American
  Mathematical Society, 1988).

\bibitem{DiVecchia:1989id}
P.~Di~Vecchia, F.~Pezzella, M.~Frau, K.~Hornfeck, A.~Lerda, and S.~Sciuto, {\it
  {N-point g-loop vertex for a free fermionic theory with arbitrary spin}},
  {\em Nucl.Phys.} {\bf B333} (1990) 635.

\bibitem{Ford1929}
L.~R. Ford, {\em {Automorphic Functions}}.
\newblock McGraw-Hill, New York, 1929.

\bibitem{Crane:1986uf}
L.~Crane and J.~M. Rabin, {\it {Super Riemann Surfaces: Uniformization and
  Teichmuller Theory}},  {\em Commun.Math.Phys.} {\bf 113} (1988) 601.

\bibitem{Lovelace:1970sj}
C.~Lovelace, {\it {M-loop generalized Veneziano formula}},  {\em Phys.Lett.}
  {\bf B32} (1970) 703--708.

\bibitem{Kaku:1970ym}
M.~Kaku and L.~Yu, {\it {The general multi-loop {V}eneziano amplitude}},  {\em
  Phys.Lett.} {\bf B33} (1970) 166--170.

\bibitem{Alessandrini:1971cz}
V.~Alessandrini, {\it {A General approach to dual multiloop diagrams}},  {\em
  Nuovo Cim.} {\bf A2} (1971) 321--352.

\bibitem{Alessandrini:1971dd}
V.~Alessandrini and D.~Amati, {\it {Properties of dual multiloop amplitudes}},
  {\em Nuovo Cim.} {\bf A4} (1971) 793--844.

\bibitem{Magnea:2013lna}
L.~Magnea, S.~Playle, R.~Russo, and S.~Sciuto, {\it {Multi-loop open string
  amplitudes and their field theory limit}},  {\em JHEP} {\bf 1309} (2013) 081,
  [\href{http://xxx.lanl.gov/abs/1305.6631}{{\tt 1305.6631}}].

\bibitem{Magnea:2015fsa}
L.~Magnea, S.~Playle, R.~Russo, and S.~Sciuto, {\it {Two-loop Yang-Mills
  diagrams from superstring amplitudes}},  {\em JHEP} {\bf 1506} (2015) 146,
  [\href{http://xxx.lanl.gov/abs/1503.05182}{{\tt 1503.05182}}].

\bibitem{Petersen:1989hi}
J.~L. Petersen, {\it {On the Explicit Construction of the Superstring Loop
  Measure in the Super-Schottky-Reggeon Formalism}},  in {\em Physics and
  Mathematics of Strings} (L.~Brink, D.~Friedan, and A.~M. Polyakov, eds.),
  pp.~434--466.
\newblock World Scientific, Singapore, 1990.

\bibitem{Maskit1967}
B.~Maskit, {\it {A characterization of Schottky groups}},  {\em J.Anal.Math.}
  {\bf 19} (1967), no.~1 227--230.

\bibitem{Giddings:1987wn}
S.~B. Giddings and P.~C. Nelson, {\it {The Geometry of Super Riemann
  Surfaces}},  {\em Commun.Math.Phys.} {\bf 116} (1988) 607.

\bibitem{Witten:2012bg}
E.~Witten, {\it {Notes On Supermanifolds and Integration}},
  \href{http://xxx.lanl.gov/abs/1209.2199}{{\tt 1209.2199}}.

\bibitem{Voronov:1987xf}
A.~A. Voronov, A.~A. Roslyi, and A.~S. Schwarz, {\it {Geometry of
  Superconformal Manifolds}},  {\em Commun. Math. Phys.} {\bf 119} (1988)
  129--152.

\bibitem{Manin1991}
Y.~I. Manin, {\em {Topics in Noncommutative Geometry}}.
\newblock Princeton University Press, Princeton, New Jersey, 1991.

\bibitem{Fairlie:1973jw}
D.~Fairlie and D.~Martin, {\it {New light on the Neveu-Schwarz model}},  {\em
  Nuovo Cim.} {\bf A18} (1973) 373--383.

\bibitem{Hornfeck:1987wt}
K.~Hornfeck, {\it {Three reggeon light cone vertex of the Neveu-Schwarz
  string}},  {\em Nucl.Phys.} {\bf B293} (1987) 189.

\bibitem{AlvarezGaume:1986es}
L.~Alvarez-Gaume, G.~W. Moore, and C.~Vafa, {\it {Theta Functions, Modular
  Invariance and Strings}},  {\em Commun.Math.Phys.} {\bf 106} (1986) 1--40.

\bibitem{Bers1975332}
L.~Bers, {\it {Automorphic Forms for Schottky Groups}},  {\em Adv.Math.} {\bf
  16} (1975), no.~3 332 -- 361.

\bibitem{Roland:1993pm}
K.~Roland, {\it {Beltrami differentials and ghost correlators in the Schottky
  parametrization}},  {\em Phys.Lett.} {\bf B312} (1993) 441--450.

\bibitem{Gardiner74}
F.~Gardiner, {\it {Automorphic forms and Eichler cohomology}},  in {\em A Crash
  Course on Kleinian Groups} (L.~Bers and I.~Kra, eds.), vol.~400 of {\em
  Lecture Notes in Mathematics}, pp.~24--47.
\newblock Springer Berlin Heidelberg, 1974.

\bibitem{Kra84}
I.~Kra, {\it {On the vanishing of and spanning sets for Poincar\'e series for
  cusp forms}},  {\em Acta Math.} {\bf 153} (1984), no.~1 47--116.

\bibitem{McIntyre:2004xs}
A.~McIntyre and L.~A. Takhtajan, {\it {Holomorphic factorization of
  determinants of laplacians on Riemann surfaces and a higher genus
  generalization of Kronecker's first limit formula}},  {\em Analysis} {\bf 16}
  (2006) 1291, [\href{http://xxx.lanl.gov/abs/math/0410294}{{\tt
  math/0410294}}].

\bibitem{D'Hoker:2015fna}
E.~D'Hoker and D.~H. Phong, {\it {Higher Order Deformations of Complex
  Structures}},  {\em SIGMA} {\bf 11} (2015) 047,
  [\href{http://xxx.lanl.gov/abs/1502.03673}{{\tt 1502.03673}}].

\bibitem{deligne}
P.~Deligne and J.~W. Morgan, {\it {N}otes on {S}upersymmetry (following
  {J}oseph {B}ernstein)},  in {\em Quantum Fields and Strings: A Course for
  Mathematicians} (P.~Deligne, P.~Etingof, D.~S. Freed, L.~C. Jeffrey,
  D.~Kazhdan, J.~W. Morgan, D.~R. Morrison, and E.~Witten, eds.).
\newblock American Mathematical Society, Rhode Island, 1999.

\bibitem{verlindeHthesis}
H.~L. {Verlinde}, {\em {The path-integral formulation of supersymmetric string
  theory}}.
\newblock PhD thesis, Utrecht university, the Netherlands, 1988.

\bibitem{Rabin:1987pe}
J.~M. Rabin, {\it {Teichmuller Deformations of Superriemann Surfaces}},  {\em
  Phys. Lett.} {\bf B190} (1987) 40.

\bibitem{Leites77}
I.~Bernshtein and D.~Leites, {\it {Integral forms and the Stokes formula on
  supermanifolds}},  {\em Funct. Anal. Appl.} {\bf 11} (1977), no.~1 45--47.

\bibitem{gunning}
R.~C. Gunning, {\em {Riemann Surfaces and Generalized Theta Functions}}.
\newblock Springer-Verlag, Berlin, 1976.

\bibitem{Mandelstam:1985ww}
S.~Mandelstam. \emph{The Interacting String Picture and Functional
  Integration}, pp.~46--102 in M.~B.~Green \emph{et al.}, eds., \emph{Unified
  String Theories} (World Scientific, 1986).

\bibitem{DiVecchia:1988cy}
P.~Di~Vecchia, F.~Pezzella, M.~Frau, K.~Hornfeck, A.~Lerda, and S.~Sciuto, {\it
  {N-point g-loop vertex for a free bosonic theory with vacuum charge Q}},
  {\em Nucl.Phys.} {\bf B322} (1989) 317.

\bibitem{Pezzella:1988jr}
F.~Pezzella, {\it {$g$ Loop Vertices for Free Fermions and Bosons}},  {\em
  Phys. Lett.} {\bf B220} (1989) 544.

\bibitem{DiVecchia:1988jy}
P.~Di~Vecchia, K.~Hornfeck, M.~Frau, A.~Lerda, and S.~Sciuto, {\it {N-string,
  g-loop vertex for the fermionic string}},  {\em Phys.Lett.} {\bf B211} (1988)
  301.

\bibitem{Cohn:1987cj}
J.~Cohn, {\it {Modular Geometry of Superconformal Field Theory}},  {\em
  Nucl.Phys.} {\bf B306} (1988) 239.

\bibitem{Giddings:1991uh}
S.~B. Giddings, {\it {Punctures on super Riemann surfaces}},  {\em
  Commun.Math.Phys.} {\bf 143} (1992) 355--370.

\bibitem{D'Hoker:2002gw}
E.~D'Hoker and D.~Phong, {\it {Lectures on two loop superstrings}},  {\em
  Conf.Proc.} {\bf C0208124} (2002) 85--123,
  [\href{http://xxx.lanl.gov/abs/hep-th/0211111}{{\tt hep-th/0211111}}].

\end{thebibliography}\endgroup
\end{document}